\documentstyle[12pt, epsf]{article}
\def\fig#1#2#3#4{
\begin{figure}
\begin{center}
\mbox{\epsfysize #1 \epsffile{#2}}
\end{center}
\caption{#3}
\label{#4}\end{figure}}

\catcode`\@=11

\@addtoreset{equation}{section}
\def\theequation{\arabic{section}.\arabic{equation}}

\def\@normalsize{\@setsize\normalsize{15pt}\xiipt\@xiipt
\abovedisplayskip 14pt plus3pt minus3pt%
\belowdisplayskip \abovedisplayskip
\abovedisplayshortskip  \z@ plus3pt%
\belowdisplayshortskip  7pt plus3.5pt minus0pt}
\def\small{\@setsize\small{13.6pt}\xipt\@xipt
\abovedisplayskip 13pt plus3pt minus3pt%
\belowdisplayskip \abovedisplayskip
\abovedisplayshortskip  \z@ plus3pt%
\belowdisplayshortskip  7pt plus3.5pt minus0pt
\def\@listi{\parsep 4.5pt plus 2pt minus 1pt
            \itemsep \parsep
            \topsep 9pt plus 3pt minus 3pt}}

\def\underline#1{\relax\ifmmode\@@underline#1\else
        $\@@underline{\hbox{#1}}$\relax\fi}
\@twosidetrue
\relax

\catcode`@=12

\catcode`\@=11
\def\lover#1{
      \raisebox{1.3ex}{\rlap{$\leftarrow$}} \raisebox{ 0ex}{$#1$}}
\def\section{\@startsection{section}{1}{\z@}{3.5ex plus 1ex minus
   .2ex}{2.3ex plus .2ex}{\large\bf}}
\def\thesection{\arabic{section}.}

\def\ps@headings{\def\@oddfoot{}\def\@evenfoot{}
\def\@oddhead{\hbox{}\hfill
        \makebox[.5\textwidth]{\raggedright\ignorespaces --\thepage{}--
        \hfill }}
\def\@evenhead{\@oddhead}
\def\subsectionmark##1{\markboth{##1}{}}
}

\ps@headings

\catcode`\@=12

\relax

\def\figcap{\section*{Figure Captions\markboth
        {FIGURECAPTIONS}{FIGURECAPTIONS}}\list
        {Fig. \arabic{enumi}:\hfill}{\settowidth\labelwidth{Fig. 999:}
        \leftmargin\labelwidth
        \advance\leftmargin\labelsep\usecounter{enumi}}}
 \relax
\def\tablecap{\section*{Table Captions\markboth
        {TABLECAPTIONS}{TABLECAPTIONS}}\list
        {Table \arabic{enumi}:\hfill}{\settowidth\labelwidth{Table 999:}
        \leftmargin\labelwidth
        \advance\leftmargin\labelsep\usecounter{enumi}}}
 \relax
\def\reflist{\section*{References\markboth
        {REFLIST}{REFLIST}}\list
        {[\arabic{enumi}]\hfill}{\settowidth\labelwidth{[999]}
        \leftmargin\labelwidth
        \advance\leftmargin\labelsep\usecounter{enumi}}}
 \relax

\catcode`\@=11

\def\ps@headings{\def\@oddfoot{}\def\@evenfoot{}
\def\@oddhead{\hbox{}\hfill
        \makebox[.5\textwidth]{\raggedright\ignorespaces --\thepage{}--
        \hfill }}
\def\@evenhead{\@oddhead}
\def\subsectionmark##1{\markboth{##1}{}}
}

\ps@headings

\relax

\def\firstpage#1#2#3#4#5#6{
\begin{document}
\begin{titlepage}
\nopagebreak
\title{\begin{flushright}
        \vspace*{-1.5in}
        {\normalsize IC/96/59 -- NUB--#1\\[-3mm]
        #2\\[-9mm] hep-th/9604077}\\[6mm]
\end{flushright}
\vfill
{\large \bf #3}}
\author{\large #4 \\[1cm] #5}
\maketitle
\begin{abstract}
{\noindent #6}
\end{abstract}
\vfill
\begin{flushleft}
\rule{16.1cm}{0.2mm}\\[-3mm]
$^{\star}${\small Research supported in part by\vspace{-4mm}
the National Science Foundation under grant
PHY--93--06906,\newline in part by the EEC contracts \vspace{-4mm}
SC1--CT92--0792 and CHRX-CT93-0340,
and in part by CNRS--NSF
grant INT--92--16146.}\\[-3mm]
$^{\dagger}${\small Laboratoire Propre du CNRS UPR A.0014.}\\
April 1996
\end{flushleft}
\thispagestyle{empty}
\end{titlepage}}     
\newcommand{\NIJ}{{\cal N}_{IJ}}
\newcommand{\N}{{\cal N}}
\newcommand{\dal}{\raisebox{0.085cm}
{\fbox{\rule{0cm}{0.07cm}\,}}}
\newcommand{\dt}{\partial_{\langle T\rangle}}
\newcommand{\dtbar}{\partial_{\langle\fracline{T}\rangle}}
\newcommand{\al}{\alpha^{\prime}}
\newcommand{\mst}{M_{\scriptscriptstyle \!S}}
\newcommand{\mpl}{M_{\scriptscriptstyle \!P}}
\newcommand{\dv}{\int{\rm d}^4x\sqrt{g}}
\newcommand{\lv}{\left\langle}
\newcommand{\rv}{\right\rangle}
\newcommand{\ph}{\varphi}
\newcommand{\sbar}{\,\overline{\! S}}
\newcommand{\xbar}{\,\overline{\! X}}
\newcommand{\barz}{\,\overline{\! Z}}
\newcommand{\zbar}{\bar{z}}
\newcommand{\dbar}{\,\overline{\!\partial}}
\newcommand{\tbar}{\overline{T}}
\newcommand{\ubar}{\overline{U}}
\newcommand{\psibar}{\overline{\Psi}}
\newcommand{\ybar}{\overline{Y}}
\newcommand{\z}{\zeta}
\newcommand{\chibar}{\bar{\chi}}
\newcommand{\dslash}{{\not\!\partial}}
\newcommand{\covslash}{\not\hspace{-0.25 em}\cal D}
\newcommand{\phb}{\overline{\varphi}}
\newcommand{\cm}{Commun.\ Math.\ Phys.~}
\newcommand{\pr}{Phys.\ Rev.\ D~}
\newcommand{\pl}{Phys.\ Lett.\ B~}
\newcommand{\ibar}{\bar{\imath}}
\newcommand{\jbar}{\bar{\jmath}}
\newcommand{\kbar}{\bar{k}}
\newcommand{\lbar}{\bar{l}}
\newcommand{\np}{Nucl.\ Phys.\ B~}
\newcommand{\e}{{\rm e}}
\newcommand{\gsi}{\,\raisebox{-0.13cm}{$\stackrel{\textstyle
>}{\textstyle\sim}$}\,}
\newcommand{\lsi}{\,\raisebox{-0.13cm}{$\stackrel{\textstyle
<}{\textstyle\sim}$}\,}
\date{}
\firstpage{3126}{CPTH--S446.0496}
{\large\sc Topological Amplitudes in Heterotic Superstring
Theory$^\star$}
{I. Antoniadis$^{\,a}$, E. Gava$^{b,c}$, K.S. Narain$^{ c}$ and
T.R. Taylor$^{\,a,d}$}
{\normalsize\sl
$^a$Centre de Physique Th\'eorique, Ecole Polytechnique,$^\dagger$
F-91128 Palaiseau, France\\[-3mm]
\normalsize\sl
$^b$Instituto Nazionale di Fisica Nucleare, sez.\ di Trieste,
Italy\\[-3mm]
\normalsize\sl $^c$International Centre for Theoretical Physics,
I-34100 Trieste, Italy\\[-3mm]
\normalsize\sl $^d$Department of Physics, Northeastern
University, Boston, MA 02115, U.S.A.}
{We show that certain  heterotic string amplitudes are given in
terms of correlators of the twisted topological (2,0)
SCFT, corresponding to the internal sector of 
the $N=1$ spacetime supersymmetric background. The genus $g$ topological
partition function $F^g$ corresponds to a term in 
the effective action of the form $W^{2g}$, where $W$ is the
gauge or gravitational superfield. We study also recursion relations 
related to holomorphic anomalies, showing
that, contrary to the type II case, they 
involve correlators of anti-chiral superfields. The corresponding terms
in the effective action are of the form $W^{2g}\Pi^n$, where
$\Pi$ is a chiral superfield obtained by chiral projection of
a general superfield. We observe that the structure of the
recursion relations is that of $N=1$ spacetime supersymmetry Ward
identity. We give also a solution of the tree level recursion
relations and discuss orbifold examples.}

\setcounter{section}{0}
\section{Introduction}

The low energy behavior of superstring theory is described by a locally
supersymmetric effective field theory of the massless modes. 
The structure of the effective theory has been studied most extensively
in the leading low-energy approximation, the so-called
two derivative level, at which the theory can be
described by the standard supergravity lagrangians.
One of the most interesting aspects of superstrings
is the relation between their low-energy interactions
and the topological properties of the underlying world-sheet $N=2$
superconformal field theory. 

In type II superstring Calabi-Yau
compactifications, special geometry of $N=2$ supergravity can be described in
the framework of a topological field theory of the twisted Calabi-Yau
sigma-model \cite{bcov}. The topological partition function of such a
model can be defined on a Riemann surface of arbitrary genus $g$. 
At genus one, it represents a new supersymmetric index \cite{index} which
determines the one-loop corrections to the four-derivative
gravitational $R^2$ couplings \cite{agn}. At higher genus, it describes a
sequence of higher derivative F-term interactions of the form
$W^{2g}$, where $W$ is the chiral superfield of $N=2$ supergravitational
multiplet \cite{agnt}. The relation between topological field
theory and four-dimensional F-terms has its origin in the topological twist
which projects on the chiral sector of the Hilbert space,
both on the world sheet and on space-time. Therefore
the topological partition function $F^g$ should be holomorphic in
chiral moduli. However, there is an anomaly of the
BRST current which induces a holomorphic anomaly in $F^g$,
captured by a set of recursion relations \cite{bcov}.
{}From the space-time point of view, this non-holomorphicity
is due to the propagation of massless states which leads to non-localities in
the effective action.

Type II superstring theory provides a natural setting for
a topological twist due to the left-right symmetry of its world sheet \cite{Top}.
Although this symmetry is absent in heterotic superstring theory,
there are indications that the F-terms of the corresponding $N=1$ 
supersymmetric effective action are also related to topological
quantities. For instance, the one loop corrections
to gauge couplings and the K\"ahler metric are still related
to the ``new" supersymmetric index \cite{agn,yuka}.
In this work, we study the heterotic version of the topological theory
obtained by twisting the left-moving (supersymmetric) sector which has
$N=2$ superconformal symmetry. The corresponding partition function
$F^g$ is related in this case to F-terms of the type $W^{2g}$, where
$W$ is now the chiral $N=1$ gauge superfield. They give rise to amplitudes
involving two gauge fields and $2g-2$ gauginos.

Unlike in the the type II case, the recursion relations describing the
holomorphic anomaly of heterotic $F^g$ couplings do not close among themselves.
Instead a new class of topological quantities appears which involve
correlation functions of anti-chiral operators. We will show that once these
quantities are included, the recursion relations close at least in the simplest
case when one considers appropriate differences between two gauge groups with no
charged massless representations. From the effective field theory point of
view, the new quantities correspond to F-terms of the type $\Pi^n W^{2g}$,
where $\Pi$'s are chiral projections of non-holomorphic functions of chiral
superfields \cite{mu}. The basic F-terms of the form $\Pi^n$ appear already
at the tree level, giving rise in particular to interactions involving
two spacetime derivatives of complex scalars and $2n{-}2$ chiral fermions.
At genus $g$, the $W^{2g}$ factor gives rise to additional $2g$
gauginos. All these terms are related by local supersymmetry
to gravitino ``mass'' terms involving $2n$ fermions and $2g$ gauginos.
Therefore they could play important role for supersymmetry breaking, in the
presence of fermion condensates induced by non-perturbative effects.

This paper is organized as follows. In section 2, we describe the heterotic
version of the topological twist and derive the corresponding partition
function related to the $W^{2g}$ F-term. In section 3, we discuss the
holomorphic anomaly and the related recursion relations. They can be put in a
form of a master equation which, at least at the tree-level, has a suggestive
interpretation as a Ward identity of $N=1$ space-time supersymmetry. In section
4, we present an algorithm for constructing solutions of the tree-level
recursion relations which close on $\Pi^n$ terms. In section 5, we exhibit
the relation of the new topological quantities appearing in the recursion
relations with physical amplitudes of the heterotic superstring theory at
arbitrary genus. In section 6, we give the superfield description of the
corresponding interaction terms as $\Pi^n W^{2g}$ F-terms. In section 7, we
work out some simple orbifold examples of the heterotic topological quantities.
In section 8, we present our conclusions and discuss the difficulties encountered
in generalizing the recursion relations to the case of gauge groups with massless
charged matter representations. Finally, in Appendices we perform explicit field
theory computations which illustrate how a tree-level $\Pi^2$ term feeds into
F-terms of the type $\Pi^3$ at the tree-level (Appendix A)
and $\Pi W^2$ at the one-loop (Appendix B), with non-holomorphic couplings
satisfying the recursion relations.

\section{Topological partition function in the heterotic case}

We begin by reviewing some of the basic features of topological field
theory obtained by twisting $N=2$ superconformal field theories that appear in
string compactifications \cite{Top}.
Let us first consider the left moving sector. One starts with an $N=2$
superconformal field theory with central charge ${\hat c}=3$ needed to
describe the compactification of the ten dimensional
superstring to four dimensions. The $N=2$ algebra is generated by
the energy momentum tensor $T$, the two supercurrents $G^+$, $G^-$,
and the $U(1)$ current $J$. Among the states of the theory, there are special
states corresponding to chiral (anti-chiral) primary fields with integral
$U(1)$ charges. These states are characterized by the relation $h=q/2$ for
chiral fields and $h=-q/2$ for the anti-chiral ones, $h$ and $q$ being their
dimensions and $U(1)$ charges, respectively. Furthermore from unitarity it
follows that $|q| \leq 3$. The states with $q=\pm 1$ give rise to 4-dimensional
chiral (anti-chiral) supermultiplets. The identity operator (namely $q=0$ state)
gives rise to the gravitational and gauge sectors of the theory.

The topological field theory is obtained
by twisting the energy momentum tensor $T\rightarrow T+{1\over 2}
\partial J$, so that the new central charge vanishes and the conformal
dimensions of the fields $h \rightarrow h-q/2$.
In particular $G^+$ acquires dimension one
and $G^-$ dimension two. Moreover, from the $N=2$
algebra, it is consistent to identify $G^+$ with the
BRST current of the topological theory $Q_{BRST}= \oint G^+$. The modified
stress energy tensor becomes  BRST-exact:
\begin{equation}
T=\{Q_{BRST}, G^- \}
\label{brst}
\end{equation}
as it should be for a topological theory. The $U(1)$ current is now
anomalous, giving rise to a background charge equal to $3(g-1)$ on
a genus $g$ Riemann surface. The chiral primaries have
dimension zero and are the physical states, as defined by
the BRST-cohomology. The anti-chiral ones become
unphysical as they are not annihilated by $Q_{BRST}$. Equation (\ref{brst})
suggests that one can identify $G^-$ with the usual reparametrization
ghost $b$. 

In the left-right symmetric theories discussed in \cite{bcov}, one
can also twist the right moving $N=2$ algebra $\bar T\rightarrow \bar T+
\frac{1}{2}\bar{\partial}\bar{J}$.\footnote{For concreteness we are considering
here type B-models with the same twist in the left- and right-moving sectors.}
$G^-$'s of left and right sectors can then be folded with holomorphic and
anti-holomorphic Beltrami differentials to define a measure over the
moduli-space of Riemann surfaces. Since the complex dimension of the moduli
space $M_g$ of genus $g$ Riemann surface is $3(g-1)$, the charges of these
$G^-$'s cancel exactly the background charge mentioned above. As a result, after
integrating over $M_g$, we get the partition
function $F^g$ of a topological string theory. 

One can also define
correlation functions of $n$ (chiral, chiral) operators of charge (1,1) each,
by inserting them at different points of the Riemann surface.
Since every insertion introduces a puncture in the surface, the dimension
of the moduli space increases by $n$ and therefore the measure involves
$n$ additional $G^-$'s and once again the background charge is balanced.
Corresponding to these zero-form operators, one has as usual two-form
operators, carrying charge zero, which can be added to the action,
representing the holomorphic moduli of the target space. In our case
they are the (1,2) (complex structure) moduli.  On the other
hand the anti-holomorphic (1,2) moduli are BRST-exact in the topological
theory, and therefore one expects that the topological correlation
functions would be holomorphic. However this naive expectation turns
out to be wrong, due to the presence of boundary terms in the integration
over $M_g$ \cite{bcov}, as we will discuss later.
(1,1) (K\"ahler) moduli (and their complex conjugates)
are also BRST-exact and can be shown to be truly decoupled.

In Refs. \cite{bcov,agnt} it was shown that certain physical amplitudes
in type II string theory corresponding to F-terms in low energy $N=2$
supergravity theory, are given by $F^g$. They involve 2 gravitons and $2(g-1)$
graviphotons. To explain why these amplitudes are related to $F^g$, let us
bosonize the left $U(1)$ current $J=i\sqrt{3}\partial H$. Then twisting of the
stress energy tensor can be accomplished by adding the term
$i\frac{\sqrt{3}}{2}
H R^{(2)}$ to the sigma model action, where $R^{(2)}$ is the 2-dimensional world
sheet curvature. Now on a genus $g$ surface, given the fact that the Euler
number is $2(1-g)$, one can choose a metric so that $R^{(2)} = -
\sum_{i=1}^{2g-2} \delta^{(2)} (x-x_i)$. This therefore suggests that the
topological partition function can be viewed as computing an amplitude in the
untwisted theory involving $2(g-1)$ vertices whose left moving part is
of the form $e^{-i\frac{\sqrt{3}}{2}H}$. Similarly bosonizing the right moving
$U(1)$ current and twisting one finds that these $2(g-1)$ vertices are
exactly the internal part of the graviphoton vertex operators in the
$-1/2$ ghost picture. Conservation of superghost charge then implies
an insertion of $3(g-1)$ picture changing operators from both left and right
sectors. By $U(1)$ charge conservation the only non vanishing terms in the
picture changing operators involve $G^-$'s giving rise to the right
number of $G^-$ insertions as is required in the definition of the topological
partition function.

To realize precisely what amplitude has to be computed, we recall
that the anti-selfdual part of the graviphoton field strength $T_{\mu\nu}$
is the lowest component of a chiral $N=2$ superfield $W_{\mu\nu}^{ij}$, from
which the scalar superfield $W^2$ can be constructed \cite{chiraln2}:
\begin{equation}
W^2 \equiv \epsilon_{ij}\epsilon_{kl}W_{\mu\nu}^{ij}W_{\mu\nu}^{kl}
=T_{\mu\nu}T_{\mu\nu}-2(\epsilon_{ij}\theta^i\sigma_{\mu\nu}
\theta^j)
R_{\mu\nu\lambda\rho}T_{\lambda\rho}-(\theta^i)^2(\theta^j)^2
R_{\mu\nu\lambda\rho}R_{\mu\nu\lambda\rho}+\dots
\label{5w2}\end{equation}
Here $R_{\mu\nu\lambda\rho}$ is the anti-selfdual part of the Riemann
tensor. It is then clear that an F-term of the type $W^{2g}$ will
contribute to an amplitude involving $2g-2$ graviphotons and $2$
gravitons.
One is thus led to compute the following amplitude:
\begin{equation}
A^g~=~\int_{M_g}\langle\int\prod_{i=1}^{2g} d^2 x_i V_h(x_1) V_h(x_2)
\prod_{i=3}^{2g}
V_T(x_i) \prod_{k=1}^{3g-3}|(\mu_k b)|^2 e^{\varphi} G^{-}(z_k)
e^{\widetilde\varphi}{\widetilde G}^{-}(\bar{z}'_k){\rangle}_g~.
\label{ampl}
\end{equation}
Here $V_T$ and $V_h$ are the vertex operators for the graviphoton and
graviton in $-1/2$ and $0$ pictures respectively. $b$ is the
reparametrization ghost, $\mu_i$'s are the Beltrami differentials,
$\varphi$ is the scalar bosonizing the superghost system and the tilde
refers to the right moving sector.

It was shown \cite{agnt} that, after using bosonization formulae and the Riemann
theta identity for summing over spin structures,
all non-zero mode determinants of ghosts, spacetime bosons and
fermions cancel, and the amplitude becomes proportional to:
\begin{equation}
A^{g}~=~\int_{M_g}\frac{1}{(\det \rm{Im}\tau)^{2}}\int\prod_{i=1}^{2g}
d^{2}x_{i} |\det{\omega_{i}(y_j)}
\det{\omega_{i}(u_{j})}|^2
\langle \prod_{k=1}^{3g-3}|(\mu_k G^{-})|^2\rangle_{\rm top},
\label{top}
\end{equation}
where $g$ $y$'s and $g$ $u$'s are partitions of the $2g$ $x$'s,
the precise partition depending on the choice of kinematics.
$\omega_i$'s are the $g$ holomorphic abelian differentials.
$(\det \rm{Im}\tau)^{-2}$ arises from the integration over spacetime
loop momenta, $\tau$ being the period matrix of the Riemann surface.
The subscript top means that the correlator is evaluated in the
internal twisted theory. The integration over $x$'s cancels then
$(\det \rm{Im}\tau)^{-2}$ and as a result we end up with just the
topological partition function.

Under K\"ahler transformations $K\rightarrow K+\phi+\bar\phi$, $\phi$
being a holomorphic function, $A^g$ transforms as $A^g\rightarrow
e^{(g-1)(\phi-\bar{\phi})}A^g$. This follows from the K\"ahler
transformation properties of the graviphoton in $N=2$ supergravity \cite{dewit},
$T_{\mu\nu}\rightarrow e^{-\frac{1}{2}(\phi-\bar{\phi})}T_{\mu\nu}$. 
Recalling that the $F^g$ transforms as $F^g\rightarrow e^{(2g-2)\phi} F^g$,
i.e.\ it has K\"ahler weight $(2g-2,0)$,\footnote{In general, for a field $\Phi$
transforming as $\Phi\rightarrow e^{(w\phi+ {\bar w}\bar{\phi})}\Phi$ the
K\"ahler weight is defined to be $(w,\bar{w})$.} we see that the $A^g$ and $F^g$
are related by $A^g=e^{(1-g)K}F^g$.

After this discussion of the type II case, it is not difficult
to guess which amplitudes in the
heterotic case are related to the heterotic version of the topological
theory. Let us recall that in the four dimensional
heterotic string with $N=1$ supersymmetry the internal theory has
an $N=2$ superconformal symmetry with $c=9$ in the left-moving sector,
while the right-moving sector has in general just conformal symmetry
with $\bar{c}=22$. Therefore one expects to have twisting only in the
left-moving sector. From the previous discussion it is then clear
that we are led to consider amplitudes involving $2g-2$ left-moving
spin fields  at genus $g$. Vertex operators having this form are
those corresponding to gauginos, gravitinos or dilatinos.
Let us first consider gauginos. An F-term giving rise to $2g-2$
gauginos can be written as $(W^{2})^{g}$, where $W^{a}_{\alpha}$ is now
the chiral gauge superfield:
\begin{equation}
W^{a}_{\alpha}=-i{\lambda}^{a}_{\alpha}-\frac{i}{2}({\sigma}^{\mu}
{\bar{\sigma}}^{\nu})_{\alpha}^{\beta}F_{\mu\nu}^{a}{\theta}_{\beta}
+\dots
\label{W}
\end{equation}
Here $a$ is the gauge group index and $\lambda$
is the gaugino field. Such an F-term would contribute to an amplitude
involving $2g-2$ gauginos and $2$ gauge fields.

The vertex operator for gauginos, in the $-1/2$ picture is:
\begin{equation}
V^\alpha_{\lambda}(p) ~=~ :e^{-\frac{1}{2} {\varphi}}
S^{\alpha}e^{i\frac{\sqrt{3}}{2}H}\bar{J}^{a} e^{ip\cdot \!X}\!:~,
\label{vgn}
\end{equation}
where $S^{\alpha}$ is spacetime spin field and $\bar{J}^{a}$ is the
right-moving Kac-Moody current.
The vertex operator for the gauge fields, in the $0$-ghost picture,
is:
\begin{equation}
V_{A}(p) ~=~ :(\partial X_{\mu} +ip\cdot\psi{\psi}_{\mu})
\bar{J}^{a} e^{ip\cdot \!X}\!:~,
\label{vgf}
\end{equation}
where $\psi$'s are spacetime fermions.
The amplitude we are interested in is therefore:
\begin{equation}
A^g~=~\int_{M_g}\langle\int\prod_{i=1}^{2g} d^2 x_i V_A(x_1) V_A(x_2)
\prod_{i=3}^{2g} V^\alpha_{\lambda}(x_i)
\prod_{k=1}^{3g-3}|(\mu_k b)|^2 e^{\varphi} G^{-}(z_k) {\rangle}_g~.
\label{hampl}
\end{equation}

Note that the left-moving part of the gauginos and gauge fields
vertices is identical to that of graviphotons and gravitons of
the type II case. Therefore, following the same steps as before,
and summing over left-moving spin structures, once again we find
that the non-zero mode part of the left-moving ghost and spacetime
boson-fermion systems cancel, and one ends up with the following
expression, up to second order in the external momenta:
\begin{eqnarray}
A^{g}~&=&~\int_{M_g}\frac{1}{(\det \rm{Im}\tau)^{2}}\int\prod_{i=1}^{2g}
d^{2}x_{i} \det{\omega_{i}(y_j)}
\det{\omega_{i}(u_{j})}\nonumber\\ & &
\langle \prod_{k=1}^{3g-3}(\mu_k G^{-})(\bar{\mu}_{k}\bar{b})
\prod_{l=1}^{g}\bar{J}^{b_l}(\bar{y}_{l})
\bar{J}^{c_l}(\bar{u}_{l}) \rangle_{\rm top}\ ,
\label{htop}
\end{eqnarray}
where, as in the type II case, $u$'s and $y$'s are partitions
of $x$'s while $b$'s and $c$'s are the gauge group indices of the currents at
positions $y$ and $u$, respectively.
The subscript top means that the correlator is
evaluated in the ``heterotic topological theory'' (HTT) , which consists of
the left moving twisted internal theory together
with the full right-moving
$\bar{c}=26$ bosonic string. Notice however that, through the factor
$(\det \rm{Im}\tau)^{-2}$, we have explicitly extracted the zero mode
contribution of the spacetime bosons $X_\mu$'s. Therefore, given the fact
that the momenta of the non-compact bosons are left-right symmetric,
the above definition of the topological theory is not strictly
rigorous. A correct definition of heterotic topological theory will
involve also the twisting of the left-moving spacetime SCFT. However
for the purposes of this paper, we shall not be needing this more 
rigorous formulation.

The physical states of the HTT are left-moving chiral operators,
corresponding to spacetime chiral superfields. The two form versions
of anti-chiral operators are BRST-exact and therefore one would expect them
to decouple. However, insertion of such operators in the correlator
(\ref{htop}) does not give a total derivative on $M_g$ due to
the presence of an explicit moduli dependent prefactor
(the $(\det \rm{Im}\tau)^{-2}$  term and the $x_i$ integrals).
Thus, to proceed further, we have to perform the $x_i$ integrals.

Let us first choose the currents along the Cartan generators
(this can always be done for small enough $g$) so that there is
no first order pole in the OPE of any two currents. The double
poles can be canceled by taking suitable differences  between
two different gauge groups. The only contributions then come from
the zero modes of the Kac-Moody currents, which are given by
$\sum_{i=1}^{g} Q_{i}^{a}{\bar{\omega}}_{i}$, where $Q_{i}^{a}$ measures
the $a$-th charge of the state propagating around the $i$-th loop. The $x_i$
integrations can then be done explicitly providing a factor $(\det
\rm{Im}\tau)^2$, and the final result is:
\begin{equation}
A^{g}~=~\int_{M_g}
\langle \prod_{k=1}^{3g-3}(\mu_k G^{-})(\bar{\mu}_{k}\bar{b})
(\det Q_{i}^{b_j}) (\det Q_{i}^{c_j})
\rangle_{\rm top}.
\label{htopf}
\end{equation}
The above expression, which is not modular invariant, is actually
just one of the terms of the modular invariant, linear combination
involving the two gauge groups which ensures the cancellation of
all double poles.

The K\"ahler transformations of $A^g$ can be deduced as before from the 
K\"ahler transformations of gauginos $\lambda\rightarrow e^{-\frac{1}{4}
(\phi-\bar{\phi})}\lambda$. Thus $A^g$ transforms as $A^g\rightarrow
e^{\frac{g-1}{2}(\phi-\bar{\phi})}A^g$. Here again one can define
$F^g=e^{\frac{(g-1)}{2}K}A^g$ which transforms holomorphically $F^g\rightarrow
e^{(g-1)\phi}F^g$, carrying K\"ahler weight $(g-1,0)$.
When discussing the recursion relations in the following section,
we will adopt, both for type II and heterotic, the prescription
of Ref. \cite{bcov} for the normalization of the vacuum state in the
topological theory, so that the topological partition function
computes directly $F^g$'s.

\section{Recursion relations}

In this section we will discuss the holomorphic anomaly and
the related recursion relations for the heterotic string.
Before doing that, it is useful to recall type II case.
Taking the antiholomorphic derivative of the topological partition
function $F^g$ amounts to the insertion of the
two form version $\Phi_{\bar{\imath}}$
of the corresponding (anti-chiral,anti-chiral) operator $\Psi_{\bar{\imath}}$.
As we have already mentioned in the previous section,
$\Phi_{\bar{\imath}}=\{Q_{BRST},\{{\bar Q}_{BRST}\Psi_{\bar{\imath}}\}\}$.
Starting from the one point function of $\Phi_{\bar{\imath}}$ and
deforming the BRST contours one gets (for $g\ge 2$):
\begin{equation}
\partial_{\bar{\imath}}F_{g}~=~\sum_{a,b=1}^{3g-3}\int_{M_g}
(-)^{(a+b)}\langle \prod_{c\ne a,d\ne b}(\mu_{c}G^{-})
(\bar\mu_{d}{\bar G}^{-})
(\mu_{a}T)(\bar\mu_{b}\bar T)\int d^2 z\Psi_{\bar{\imath}}(z,\bar{z})\rangle\ .
\label{anom}
\end{equation}
The insertions of the energy momentum tensors $T$ and ${\bar T}$ give rise
to total derivatives in the moduli space $M_g$. It follows that possible
contributions to $\partial_{\bar{\imath}} F^{g}$ 
arise from the boundaries of $M_g$, i.e.\ from Riemann surfaces with nodes. 

The analysis of the boundary contributions 
is very similar in the two cases of pinching a handle
or a dividing geodesic: parameterize the opening of a node by the complex
coordinate $t$, then due to the fact that in (\ref{anom}) one has
${\partial_t}\partial_{\bar t}$
acting on the one point function of $\Psi_{\bar{\imath}}$
on a surface with node, the only non vanishing contribution comes when
this one point function behaves like $\ln(t\bar t)$ as $t\rightarrow 0$.
This behavior appears when the operator $\Psi_{\bar{\imath}}$ sits in the node
and interacts with two anti-chiral operators via Yukawa coupling
$C_{\bar{\imath}\bar{\jmath}\bar{k}}$,
the integration of its position giving $\ln (t\bar t)$. Finally the two
anti-chiral operators propagate to chiral ones giving rise to
holomorphic derivatives of lower genus partition function. Thus
one gets the following recursion relation:
\begin{equation}
\partial_{\bar{\imath}} F^g = \frac{1}{2} C_{\bar{\imath}\bar{\jmath}\bar{k}}
e^{2K} G^{i\bar{\imath}}
G^{j\bar{\jmath}} [~\sum_{g'=1}^{g-1} D_{j} F^{g'} D_k F^{g-g'}
+ D_{j} D_k F^{g-1}~]
\label{anomt}
\end{equation}
Here $K$ is the K\"ahler potential and $G^{i\bar{\imath}}$ is the
inverse of the K\"ahler metric and $D$ is the K\"ahler covariant
derivative. Since $F^g$ transforms as $F^{g}\rightarrow
e^{(2g-2)\phi}F^{g}$ under the K\"ahler transformations,
we have  $D_{i}F^{g}\equiv ({\partial_{i}}+(2g-2)K_{i})
F^{g}$.\footnote{In general, on a quantity of K\"ahler weight $(w,\bar{w})$
the K\"ahler covariant derivatives are $({\partial_{i}}+w K_{i})$ 
and $({\partial_{\ibar}}+\bar{w} K_{\ibar})$, respectively.}  
The crucial point here is that due to the presence of
${\partial_t}\partial_{\bar t}$, the extra operator $\Psi_{\bar{\imath}}$
gets stuck in the node and as a result what appears in the right
hand side of (\ref{anomt}) again involves holomorphic derivatives
of partition functions. In other words recursion relations
close within the set of all partition functions. As we shall see
below in heterotic case this does not happen.

Indeed let us take a derivative of $F^g$ (related to the
$A^g$ of (\ref{htopf}) as explained in the previous section) with respect to
some anti-holomorphic modulus. The two form $\Phi_{\bar{\imath}}$
corresponding to this modulus is again BRST-exact,
$\Phi_{\bar{\imath}}=\{Q_{BRST},\Psi_{\bar{\imath}}\}$, where $\Psi_{\ibar}$
is dimension $(1,1)$ anti-chiral operator with charge (-1). Note that now only
one BRST (left-moving) operator appears in this expression.
Inserting $\Phi_{\bar{\imath}}$ in $F^g$ and deforming
the BRST-contour one finds:
\begin{equation}
\partial_{\bar{\imath}}F^{g}~=~\int_{M_g}\sum_{m=1}^{3g-3} (-1)^m
\langle (\mu_m T)\prod_{n\ne m} (\mu_n G^-)
\prod_{k=1}^{3g-3} (\bar{\mu}_k \bar b)\int \Psi_{\bar{\imath}}
(\det Q_{i}^{b_j}) (\det Q_{i}^{c_j})
\rangle_{\rm top}.
\label{anom1}
\end{equation}
Once again we get a total derivative in the world-sheet
moduli space and the contribution coming from the boundaries
corresponding to Riemann surfaces with nodes is
$\partial_t$ of the one point function of $\Psi_{\bar{\imath}}$.
The integral over $t,\bar t$ in the limit $t\rightarrow 0$
will be non-vanishing only if this one-point function behaves as
$1/\bar t$. This indicates that the intermediate state going through
the node is a chiral massless state (i.e. has left and
right-moving dimensions equal to 0 and 1 respectively) and moreover the 
operator $\Psi_{\bar{\imath}}$ should be distributed
over the complement of the node.\footnote{In fact if there is a non-zero
contribution when $\Psi_{\bar{\imath}}$ is at the node then this would
result in a logarithmic divergence due to the appearance of
$\ln (t\bar t)$. This could occur if there are massless states in the
theory which acquire a mass in the presence of non-zero vacuum
expectation value for the $\bar{\imath}$- modulus. In the following we
assume that we are dealing with a region of moduli space where
massless states remain massless.} What are the possible intermediate
states? 

Let us first consider the case of pinching a dividing geodesic,
so that the surface splits into two components $\Sigma_1$ and $\Sigma_2$
of genus $g_1$ and $g_2$ ($g_1 +g_2 =g$)
with punctures $P_1$ and $P_2$, respectively.
The number of $G^-$'s as well as $\bar b$'s
on $\Sigma_1$ and $\Sigma_2$ is $(3g_1-3+1)$ and
$(3g_2 -3 +1)$, respectively, due to the fact that there is one extra
world-sheet modulus for each puncture.
We can assume that the operator $\Psi_{\bar{\imath}}$ is on $\Sigma_1$.
Since the operator $\Psi_{\bar{\imath}}$ carries $U(1)$ charge $(-1)$, the
anomalous conservation of the $U(1)$ charge implies that
the intermediate state at the puncture $P_1$ must have a charge (+2).
This further implies that the state at $P_2$ being dual to the one
at $P_1$ must have a charge (+1) which is also consistent with the
$U(1)$ anomaly on $\Sigma_2$. The right moving part of the intermediate
state must be gauge group singlet. On $\Sigma_2$ using the extra $G^-$,
we can convert the charge (+1) state into neutral two-form operator which
just corresponds to taking a holomorphic K\"ahler 
covariant derivative $D_{j}$
of $F^{g_2}$. Note that here $j$ labels any chiral
gauge singlet and not necessarily only moduli. On $\Sigma_1$ the charge
(+2) state of dimension $(0,0)$
appearing at the puncture can be identified with the state
obtained by the action of the holomorphic three form $\rho$ on an
anti-chiral charge (-1) state $\bar{c}\Psi_{\bar{\jmath}}$:
$\oint dz \rho (z) \bar{c}\Psi_{\bar{\jmath}}$. Here $\bar{c}$ is the
right-moving reparametrization 
ghost.\footnote{Recall that $\rho$ is a chiral charge (+3) operator with
dimension 0 and exists for all internal theories leading to space-time
supersymmetry.} Thus on $\Sigma_1$ we end up with a new object 
$F^{g_1}_{\bar{\imath};\bar{\jmath}}$
that is a two point function involving an unphysical charge (-1)
operator $\Psi_{\bar{\imath}}$ and the charge (+2) operator
$\oint \rho \Psi_{\bar{\jmath}}$. Therefore the holomorphic anomaly
is given by:
\begin{equation}
\partial_{\ibar} F^g~=~ \sum_{g_1 + g_2 = g} F^{g_1}_{\ibar;\jbar} 
G^{\jbar j}
D_{j} F^{g_2} +\cdots
\label{danom}
\end{equation}
where $g_1,g_2>0$ and the dots refer to contributions coming from
pinching a handle. Notice that in order for this equation to 
make sense, $F^{g_1}_{\ibar;\jbar}$ must carry K\"ahler weight
$(g_1,0)$, which indeed turns out to be the case as we shall see below.

In the case of pinching a handle, we expect a contribution from
a configuration in which a $(Q^a)^2$ is inserted on the handle itself.
Thus, one obtains a two-point function on genus $g-1$ Riemann surface 
involving a massless state charged under the gauge group.
As in the dividing case, after the pinching there are $3g-4$ $G^-$'s,
two of which are used to convert the states at the punctures in their
integrated two-form version, while the remaining $3g-6$ are used
to define the genus $g-1$ two-point function.
To simplify the discussion, we will consider models with
pure gauge groups without charged matter fields. In this case the handle
contribution vanishes for the following reason: the two conjugate
states propagating in the handle
involve the identity operator and its conjugate three-form $\rho$
in the internal $N=2$ superconformal theory, therefore acting with $G^-$
on the identity we get zero. In section 7, we discuss
explicit construction of such models, while
in the last section we comment on difficulties concerning the
generalization to the case with charged matter fields.

Unlike the type II case,
the appearance of new objects besides $F^g$'s on the right hand
side of equation (\ref{danom}) shows that the recursion relations do not
close within the topological correlation functions. Indeed from the
definition of $F^{g_1}_{\ibar;\jbar}$
it is clear that we have to allow operators in the twisted theory that
are not in the kernel of $Q_{BRST}$. Nevertheless we shall see below 
that these
quantities are also related to physical amplitudes in heterotic string theory.
One can ask the question whether one can obtain some recursion relations
for  $F^{g_1}_{\ibar;\jbar}$. To see this let us take an anti-holomorphic
derivative of $F^{g_1}_{\ibar;\jbar}$. This amounts to insertion of
$\Phi_{\bar{k}}=\{Q_{BRST},\Psi_{\bar{k}}\}$ in this two point function.
This time however the deformation of BRST-contour does not give only
boundary terms due to the fact that there is a contribution when
$Q_{BRST}$ acts on $\Psi_{\ibar}$. Note that $Q_{BRST}$ annihilates
$\oint \rho \Psi_{\bar{\jmath}}$ as it is expected from the fact that the
latter is a charge (+2) chiral operator. If one anti-symmetrizes $\bar{k}$
and $\ibar$ then only the boundary terms survive, i.e. those coming
from the action of $Q_{BRST}$ on the $G^-$'s. One can then repeat the 
analysis
which led to (\ref{danom}) and the result from pinching a dividing geodesic is:
\begin{equation}
D_{[\bar{k}} F^g_{\ibar];\jbar} = \sum_{g_1+g_2=g}G^{m \bar m}
( F^{g_1}_{\bar k \ibar;\jbar \bar m} D_m F^{g_2} +
F^{g_1}_{[\bar k; \bar m} D_m F^{g_2}_{\ibar];\jbar})  \ ,
\label{danom1}
\end{equation}
where square brackets denote anti-symmetrization of $\bar{k}$ and $\ibar$.
Notice that on the right hand side of (\ref{danom1}) there appears yet
another new object with four indices $F^{g_1}_{\bar k \ibar;\jbar \bar m}$
which is defined as the 4-point function involving $\Psi_{\bar k}$,
$\Psi_{\ibar}$, $\oint dz \rho(z)\bar{c}\Psi_{\jbar}$ and $\oint dz\rho(z)
\bar{c} \Psi_{\bar m}$.
Since $\Psi$'s are anti-commuting this 4-point function is anti-symmetric
in $\bar k$ and $\ibar$, and similarly due to the presence of $\bar{c}$
it is also anti-symmetric in $\jbar$ and $\bar m$.

It is clear now that repeating this procedure one would keep getting
higher and higher point functions. The general structure of these terms is
\begin{eqnarray}
F^{g}_{\ibar_1 \cdots \ibar_n~;~\jbar_1 \cdots \jbar_n} &=& \int_{M_{g,n}}
\langle \prod_{k=1}^{3g-3+n}(\mu_k G^{-})(\bar{\mu}_{k}\bar{b})
(\det Q_{i}^{b_j}) (\det Q_{i}^{c_j})
\nonumber \\ & &\hskip 1cm
\int \Psi_{\ibar_1}\cdots \int \Psi_{\ibar_n} \widetilde{\Psi}_{\jbar_1} \cdots
\widetilde{\Psi}_{\jbar_n}
\rangle_{\rm{top}}.
\label{nptf}
\end{eqnarray}
where $\widetilde{\Psi}_{\jbar} = \oint dz \rho(z) \bar{c} \Psi_{\jbar}$.
Here $M_{g,n}$ is the moduli space of genus $g$ Riemann surface with $n$
punctures corresponding to the insertion of  $\widetilde{\Psi}_{\jbar}$'s.
Note that $\Psi_{\ibar}$'s being of dimension $(1,1)$ are integrated over the
surface.
As in the $n=2$ case, $F^{g}_{\ibar_1 \cdots \ibar_n~;~\jbar_1 \cdots \jbar_n}$
is totally antisymmetric in the $\ibar$ and $\jbar$ indices.
Moreover one can prove the following identity:
\begin{equation}
F^{g}_{[\ibar_1 \cdots \ibar_n~;~\jbar_1 ]\jbar_2 \cdots \jbar_n}~=~0.
\label{ident}
\end{equation}
Here the square bracket indicates total antisymmetrization of all
the enclosed indices $[\ibar_1 \cdots \ibar_n~;~\jbar_1]$.
To see this, convert the $\widetilde{\Psi}_{\jbar_1}$ operator
into its two form version by acting with $G^-$ and $\bar b$: 
\begin{equation}
\oint dz G^{-}(z)\oint d{\bar z}{\bar b}({\bar z})
\widetilde{\Psi}_{\jbar}=\oint dz {\cal J}(z) \Psi_{\jbar_1},
\label{J++}
\end{equation}
where ${\cal J}=\oint dz G^{-}(z)\rho$ is a charge $(+2)$ bosonic current,  
which in the twisted theory has dimension $1$. We can now substitute
eq.(\ref{J++}) into (\ref{nptf}) and deform the contour
of $\cal J$. Using the fact that it annihilates
$\widetilde{\Psi}$ and also the antisymmetry properties of $\Psi$'s
one immediately gets the identity (\ref{ident}).
 
Let us now turn to the question whether these higher point functions
in the twisted theory have a counterpart in the heterotic string theory.
As we have already seen the partition function corresponds to a string
amplitude involving 2 gauge fields and $(2g-2)$ gauginos. By the internal
$U(1)$ conservation as well as superghost charge conservation this amplitude
involves $3g-3$ $G^-$'s as is required for topological partition function.
For the $2n$-point function above we have $3g-3+n$ $G^-$'s. First of all
since these insertions correspond to anti-chiral operators we must
consider a string amplitude involving $2n$ additional anti-chiral fields.
We will insert $2n$ anti-chiral fermions $\bar{\chi}_{\ibar}$
whose vertex operators in the $-1/2$ ghost picture are given by:
\begin{equation}
V^{\dot{\alpha}}_{\ibar} =
e^{-\frac{\varphi}{2}} S^{\dot{\alpha}}\Sigma_{\ibar}e^{ip\cdot X},
\label{fvertex}
\end{equation}
where $\Sigma_{\ibar} = \oint dz \frac{1}{\sqrt{z}} e^{i\frac{\sqrt{3}}{2} H(z)}
\Psi_{\ibar}$  carries $U(1)$ charge $(+1/2)$. This necessitates the presence
of $n$ additional picture changing operators. From
$U(1)$ charge conservation it follows that only the part of the
picture changing operators containing $G^-$ contributes to the amplitude.

We are thus led to consider a genus $g$ string amplitude of the form
$\langle F_{\mu\nu}^2 \lambda^{2g-2} \bar{\chi}^{2n} \rangle$. In
section 5, we show that indeed, after using bosonization formulae and summing
over spin structures, this amplitude is proportional
to $e^{\frac{1}{2}(g-1+n)K}F^{g}_{\ibar_1 \cdots \ibar_n~;
~\jbar_1 \cdots \jbar_n}$, the
two types of indices $\ibar$ and $\jbar$ corresponding to the two
spin fields $S^{\dot{1}}$ and and $S^{\dot{2}}$. 
We shall also see in section 6
that these amplitudes are related to some higher weight F-terms
in the supergravity action. From this identification it also
follows that the K\"ahler weight of 
$F^{g}_{\ibar_1 \cdots \ibar_n~;~\jbar_1 \cdots \jbar_n}$
is $(g-1+n,0)$, since the $\bar\chi$'s transform as $\lambda$'s.

The arguments leading to (\ref{danom1}) can now be used to obtain
the recursion relation for $F^{g}_{\ibar_1 \cdots \ibar_n~;~\jbar_1
\cdots \jbar_n}$, with the result
\begin{equation}
D_{[\ibar} F^{g}_{\ibar_1 \cdots \ibar_n]~;~\jbar_1 \cdots \jbar_n}
~=~ \sum_{g_1+g_2=g} \sum_{m=0}^{n} \sum_{\rm{partitions}}
F^{g_1}_{\bar{k}_1\cdots \bar{k}_{m+1}~;~\bar{\ell}_1\cdots \bar{\ell}_m
\bar{\ell}} G^{\ell \bar{\ell}} D_{\ell} F^{g_2}_{\bar{k}_{m+2}\cdots
\bar{k}_{n+1}~;~\bar{\ell}_{m+1}\cdots \bar{\ell}_n}  \ .
\label{hanomn}
\end{equation}
Here $\bar{k}_1 \cdots \bar{k}_{n+1}$ are different permutations
of $\ibar, \ibar_1 \cdots \ibar_n$ while $\bar{\ell}_1 \cdots \bar{\ell}_n$
are permutations of $\jbar_1 \cdots \jbar_n$. The sum over permutations
is weighted by their respective signs.
For $g_1=0$ the range of $m$ is between 1 and n while
for $g_2=0$ the range is between 0 and $(n-2)$. 
Notice that this equation is consistent with the K\"ahler weight
assignments we have given above.

If one considers $F^{g}_{\ibar_1 \cdots \ibar_n~;~\jbar_1 \cdots \jbar_n}$
as a differential form in the $\ibar$ indices, then eq.(\ref{hanomn})
is a statement about its non-closedness.
This equation can be further put in a succinct form in terms of
a generating function. Recalling that (\ref{nptf}) is totally
anti-symmetric in $\ibar$ and $\jbar$ indices separately, we
introduce Grassman parameters $\theta^{\ibar}$ and $\eta^{\jbar}$.
One can now define the generating function as:
\begin{equation}
F^g = -\sum_{n \ge 2}\frac{1}{(n!)^2} \theta^{\ibar_1}\eta^{\jbar_1} \cdots
\theta^{\ibar_n}\eta^{\jbar_n}
F^{g}_{\ibar_1 \cdots \ibar_n ~;~\jbar_1 \cdots \jbar_n}
\label{genf}
\end{equation}
The identity (\ref{ident}) in terms of the generating
function $F^g$ becomes:
\begin{equation}
\theta^{\ibar}\frac{\partial F^g}{\partial \eta^{\ibar}}=0.
\label{ident1}
\end{equation}
Finally the recursion relation (\ref{hanomn}) can then be recast as:
\begin{equation}
\theta^{\ibar} D_{\ibar} F^g = \sum_{g_1+g_2=g}\frac{\partial 
F^{g_1}}{\partial \eta^{\jbar}} G^{j \jbar} D_{j} F^{g_2}.
\label{recg1}
\end{equation}
Here it is understood that the covariant derivatives act just
on the coefficients 
$F^{g}_{\ibar_1 \cdots \ibar_n ~;~\jbar_1 \cdots \jbar_n}$, and not
on $\theta$'s and $\eta$'s. In other words  $D_{\ibar}F^{g}=
\partial_{\ibar}F^{g}-\Gamma_{\ibar \jbar}^{\bar{k}}(\theta^{\jbar}
\frac{\partial}{\partial \theta^{\bar{k}}} + \eta^{\jbar}
\frac{\partial}{\partial \eta^{\bar{k}}})F^{g}$
and $D_{i} F^g= \partial_{i}F^g + K_{i} 
(\eta^{\jbar}
\frac{\partial}{\partial \eta^{\jbar}}-1)F^{g}$.

An important consistency condition for (\ref{recg1}) is provided
by the integrability condition 
$(\theta^{\ibar} D_{\ibar})^2 F^g =0$. Indeed one can verify
by using antisymmetry properties of Grassman variables, equation 
(\ref{recg1}) and the identity (\ref{ident1}) that the operator 
$\theta^{\ibar} D_{\ibar}$ annihilates the right hand side of
(\ref{recg1}).

Specializing eq.(\ref{hanomn}) to the case $g=0$ (where there 
is obviously no handle degeneration in any model), the
anomaly equation becomes:
\begin{equation}
\theta^{\ibar} D_{\ibar} F^0 = \frac{\partial 
F^{0}}{\partial \eta^{\jbar}} G^{j \jbar} D_{j} F^{0}.
\label{recg}
\end{equation}

Equation (\ref{recg}) is strongly suggestive of space-time 
supersymmetry Ward identity. Consider the anti-chiral super
multiplet $(z^{\ibar}, \bar{\chi}^{\ibar}, h^{\ibar})$ where
$h^{\ibar}$ is the auxiliary field and $\bar{\chi}^{\ibar}$ is
a two component spinor with the two components being 
$\theta^{\ibar}$ and $\eta^{\ibar}$. One of the supersymmetry
transformations on these fields is:
\begin{eqnarray}
\delta_s z^{\ibar} &=& \theta^{\ibar} \nonumber \\
\delta_s \eta^{\ibar} &=& - h^{\ibar} \nonumber \\
\delta_s \theta^{\ibar} &=& \delta_s h^{\ibar} = 0,
\label{susy}
\end{eqnarray}
and the variation of all holomorphic fields is zero. Then the
Ward identity for such a transformation on the effective action $S$ is:
\begin{equation} 
\theta^{\ibar} \partial_{\ibar} S = - h^{\ibar} \frac{\partial}
{\partial \eta^{\ibar}} S
\label{ward}
\end{equation}
If now we identify $S$ with $F^0$ and $-h^{\ibar}$ with
$(\Gamma_{\jbar \bar{k}}^{\ibar} \theta^{\jbar} \eta
^{\bar{k}} + G^{i\ibar} D_{i} F^0)$ then we obtain
(\ref{recg}). The above argument was heuristic, however in the following 
section where we describe a method to construct a solution of the
recursion relations, we shall make this argument more precise. It is also
interesting to note that a similar result has been obtained recently
\cite{deWit2} for the $N=2$ recursion relation (\ref{}). The first term on the
r.h.s. which originates from handle degeneration has been found as a consequence
of $N=2$ space-time supersymmetry.

\section{Solution to the tree level recursion relations}

We now present a topological Feynman diagrammatic solution to the recursion
relation (\ref{recg}). The first point to note is that there is an 
ambiguity in the solution. Since eq.(\ref{recg}) (or (\ref{hanomn})
with $g=0$)
relates $n$-forms\footnote{The 
degree of the form here refers to the indices that are 
contracted with $\theta^{\ibar}$'s} on the 
left hand side to lower forms on the right hand side, at each step one can 
always add a closed $n$-form to a given solution and get another solution. 
This is analogous to the holomorphic ambiguity in the type II case. In 
fact these closed $n$-forms are associated to the true 
F-terms in the effective action, to be discussed in 
section 6.
They represent contributions of massive string modes as contrasted to
the non-closed part which is generated by connected graphs
with vertices corresponding to lower forms and massless propagators.
Our solution will be obtained by starting with a set of 
closed $n$-forms ($n\ge2$), which can be thought of as genuine F-terms
from the effective field theory point of view. Thus consider the $n$-form
\begin{equation}
F^{[n]} = F^0_{\ibar_1,\dots,\ibar_n ; \jbar_1, \dots,\jbar_n}\theta^{\ibar_1}
\cdots \theta^{\ibar_n} \eta^{\jbar_1}\cdots \eta^{\jbar_n}
\label{fndef}
\end{equation}
which satisfies:
\begin{equation}
\delta F^{[n]} \equiv \theta^{\ibar}\partial_{\ibar} F^{[n]}=0, 
\qquad\qquad
\theta^{\ibar}\frac{\partial F^{[n]}}{\partial \eta^{\ibar}} =0
\label{f0n}
\end{equation}
Under 
K\"ahler transformation we assign weight $(-\frac{1}{2},0)$ to
$\theta$'s and $\eta$'s, so that $F^{[n]}$ have weight $(-1,0)$.

We would now like to construct the vertices corresponding to $F^{[n]}$.
The set of vertices can be grouped together in the following $n$-th order
polynomial in the $M+1$ variables $h^{\bar{A}}$, $A=0$ or $i$ 
with $i=1,\dots,M$ where $M$ is the number of matter fields:
\begin{equation}
\tilde{F}^{[n]} = F^{[n]} + \sum_{k=1}^n 
F^{(n,k)}_{\bar{A}_1\dots\bar{A}_k}h^{\bar{A}_1}\dots h^{\bar{A}_k}
\label{tildefn}
\end{equation}
with the coefficients $F^{(n,k)}$ being functions of 
$z^i,z^{\ibar},\theta^{\ibar},
\eta^{\ibar}$. We require that $\tilde{F}^{[n]}$ satisfy the equations:
\begin{eqnarray}
\delta \tilde{F}^{[n]} &=& h^{\ibar}\frac{\partial \tilde{F}^{[n]}}
{\partial \eta^{\ibar}} + \theta^{\ibar} h^{\bar{0}}\frac{\partial
\tilde{F}^{[n]}} {\partial h^{\ibar}},\nonumber\\
\theta^{\ibar}\frac{\partial \tilde{F}^{[n]}}{\partial \eta^{\ibar}}&=&0.
\label{tilfneq}
\end{eqnarray}
It is clear from these equations that $F^{(n,k)}$ contain $(n-k)$ 
$\theta^{\ibar}$'s and $\eta^{\ibar}$'s each. Note that the above 
equations are in fact a statement of supersymmetry invariance of 
$\tilde{F}^{[n]}$'s, where we define supersymmetry transformations on the
variables as:
\begin{equation}
\delta_s z^{\ibar} = \theta^{\ibar}, ~~~~
\delta_s \eta^{\ibar} = -h^{\ibar}, ~~~~ 
\delta_s h^{\ibar} = -\theta^{\ibar} h^{\bar{0}},~~~~
\delta_s z^i = \delta_s \theta^{\ibar} = \delta_s h^{\bar{0}} = 0
\label{susys}
\end{equation}
It is clear that first of the equations (\ref{tilfneq}) is just the 
statement that $\tilde{F}^{[n]}$ is invariant under this 
transformation. The variation  $\delta_s$ is 
just half of the supersymmetry transformation that act on anti-chiral 
fields and the invariance of $\tilde{F}^{[n]}$ has the significance that 
it represents the lowest component of a chiral superfield. Note that its action
on the fields (\ref{susys}) is not nilpotent. In section 6, we will discuss the
relation of these transformations to the usual supersymmetry transformations in
$N=1$ Poincar\'e supergravity.

Under analytic 
reparametrizations: $z^i \rightarrow 
z'^i(z)$, we demand that $\tilde{F}^{[n]}$ be invariant provided 
\begin{equation}
h^{\ibar} \rightarrow \frac
{\partial z'^{\ibar}}{\partial z^{\jbar}} h^{\jbar}
+\frac{\partial^2 z'^{\ibar}}{\partial z^{\jbar}\partial z^{\bar{k}}}
\theta^{\jbar} \eta^{\bar{k}}.
\label{phibart}
\end{equation}
Of course $\theta$ and $\eta$ transform as vectors. Under  K\"ahler
transformations we assign weight $(-1,0)$ to $h^{\bar{A}}$'s, as required by
the consistency of equations  (\ref{tilfneq}) and (\ref{phibart}) and as a
result $\tilde{F}^{[n]}$ have also  weight $(-1,0)$. 

For $n=1$ we define 
\begin{equation}
\tilde{F}^{[1]} = \theta^{\ibar} \eta^{\jbar} K_{\ibar \jbar} -
h^{\ibar} K_{\ibar} - h^{\bar{0}} K
\label{f1}
\end{equation}
where $K$ is the K\"ahler potential and subscripts on $K$ denote partial
derivatives.
One can easily see that $\tilde{F}^{[1]}$ satisfies the conditions
(\ref{tilfneq}) and (\ref{phibart}).
The action is then defined as:
\begin{equation}
S= \sum_{n\geq 1} e^{(n-1)(z^0+\phi^0)} \tilde{F}^{[n]} (z^i + \phi^i, 
z^{\ibar}, \theta^{\ibar}, \eta^{\ibar}, h^{\bar{A}}) -
\tilde{F}^{[1]} (z^i, z^{\ibar},
\theta^{\ibar}, \eta^{\ibar}, h^{\bar{A}}) - \phi^0 h^{\bar{0}}
\label{action}
\end{equation}
We regard $\phi^A$ and $h^{\bar{A}}$ as quantum variables which are to
be integrated out. The variable $z^0$ above is just introduced for 
convenience and we shall set it to zero at the end. The action $S$ is 
invariant under analytic reparametrization
with $h^{\ibar}$ transforming as (\ref{phibart}) and $\phi^i$ 
transforming as $\phi^i \rightarrow z'^i (z+\phi) - z'^i(z)$. 
{}From the foregoing discussion it is clear that under the K\"ahler
transformation $K\rightarrow K+f(z)+\bar{f}(\bar{z})$, 
\begin{equation}
\tilde{F}^{[n]}(z+\phi)
\rightarrow e^{(n-1)f(z+\phi) -nf(z)} (\tilde{F}^{[n]}(z+\phi) -
\delta_{n,1} h^{\bar{0}} (f(z+\phi) + \bar{f}(\bar{z}))
\label{fnkahler}
\end{equation}
The action $S$ then transforms as weight $(-1,0)$ provided
$\phi^0 \rightarrow \phi^0 - f(z+\phi) + f(z)$.

Consider the tree level effective action $W$ which is
obtained by substituting the classical solution of $\phi^A$ and 
$h^{\bar{A}}$ in the action $S$. The effective action $W$ can be 
expanded as:
\begin{equation}
W = -\sum_{n\geq 1} e^{(n-1)z^0} W^{[n]}_{\ibar_1,\dots,\ibar_n ; \jbar_1, 
\dots,\jbar_n}\theta^{\ibar_1}\eta^{\jbar_1}
\cdots \theta^{\ibar_n}\eta^{\jbar_n}
\label{wn}
\end{equation}
where $W^{[n]}$ depend on $z^i$ and $z^{\ibar}$'s. We have extracted 
above the explicit dependence on $z^0$ by using the scaling property of the
action $S$ under the rescalings: 
\begin{equation}
\theta^{\ibar} \rightarrow e^{\alpha} \theta^{\ibar}, ~~~~\eta^{\ibar}
\rightarrow e^{\alpha}\eta^{\ibar}, ~~~~h^{\ibar}\rightarrow e^{2\alpha} 
h^{\ibar}, ~~~~ z^0 \rightarrow z^0 - 2\alpha
\label{scale}
\end{equation}  
One can actually show, using the form of the action $S$, that $n=1$ term in 
the above expansion vanishes.
{}From the transformation properties of $S$ discussed above, it follows that
the coefficients $W^{[n]}$ transform covariantly under analytic 
reparametrizations and have K\"ahler weight $(n-1,0)$

We now show that $W$ satisfies
the recursion relation (\ref{recg}) where reparametrization and K\"ahler 
covariant derivatives appear. 
We start by applying $\delta$ on $W$. Since $W$ is the tree level 
effective action it follows that: 
\begin{equation}
\delta W = (\delta + (\delta \phi^A_{\rm 
cl})\frac{\partial}{\partial \phi^A} + (\delta h^{\bar{A}}_{\rm 
cl})\frac{\partial}{\partial h^{\bar{A}}}) S|_c = \delta S|_c
\label{deltaw}
\end{equation}
where $\phi_{\rm cl}$ denotes the classical solution and the 
symbol $|_c$ means that the corresponding quantity is evaluated at 
the classical solution. In the second step we have used the equation of 
motion. Using then (\ref{tilfneq}) and once again using the equation of 
motion we get: 
\begin{equation}
\delta W =  h^{\ibar}_{\rm cl} \frac{\partial W}{\partial
\eta^{\ibar}} 
\label{deltas}
\end{equation}
To find the classical solution for $h^{\ibar}$, we vary $S$ with 
respect to $\phi^i$ and obtain 
\begin{equation}
h^{\ibar}_{\rm cl} = G^{i\ibar} \frac{\partial W}{\partial z^i}  
+\theta^{\jbar} \eta^{\bar{k}} \Gamma_{\jbar \bar{k}}^{\ibar} - G^{i\ibar}
h^{\bar{0}}_{\rm cl} K_i
\label{phibars}
\end{equation}
The second term on the right hand side above when substituted in 
(\ref{deltas}) covariantizes the $\delta$ appearing on the left hand
side of (\ref{deltas}). Finally the classical solution 
for $h^{\bar{0}}$ is obtained 
by varying the action with respect to $\phi^0$ with the result
\begin{equation}
h^{\bar{0}}_{\rm cl} = \frac{\partial W}{\partial z^0}
= (\eta^{\ibar}\frac{\partial}{\partial \eta^{\ibar}} - 1)W
\label{phi0}
\end{equation}
where in the second equality we have used the fact that $W$ is of the 
form given in (\ref{wn}).
This expression when substituted in (\ref{phibars}) K\"ahler covariantizes
the derivative with respect to $z^i$ appearing on the right hand side of
(\ref{phibars}). Thus combining equations (\ref{deltas}), (\ref{phibars}) and
(\ref{phi0}) one finds that $W$ satisfies the recursion relation (\ref{recg}).

Now we construct $\tilde{F}^n$ explicitly for the case when $F^n$ is
of the form described in section arising from $\Pi^n$ term,
namely
\begin{equation}
F^{[n]} = f^1_{\ibar_1 \jbar_1}\theta^{\ibar_1}\eta^{\jbar_1}\cdots f^n_
{\ibar_n \jbar_n}\theta^{\ibar_n}
\eta^{\jbar_n}
\label{fnpin}
\end{equation}
Then the following $\tilde{F}^n$ satisfies the conditions (\ref{tilfneq}):
\begin{equation}
\tilde{F}^{[n]} = \sum_{k=0}^n \frac{1}{k!(n-k)!} \sum_{\sigma}
f^{\sigma(1)}_{\ibar_1 \jbar_1}\theta^{\ibar_1}\eta^{\jbar_1}\cdots f^{
\sigma(k)}_
{\ibar_k \jbar_k} \theta^{\ibar_k}\eta^{\jbar_k} 
f^{\sigma(k+1)}_{\bar{A}_1}h^{\bar{A}_1}\cdots 
f^{\sigma(n)}_{\bar{A}_{n-k}} h^{\bar{A}_{n-k}}
\label{tilfnpin}
\end{equation}
where $f^a_{\bar{0}} \equiv f^a$ and $\sigma$ denotes an element of the
permutation group of $n$-objects. In fact this expression for 
$\tilde{F}^{[n]}$
is the lowest component of the superpotential term with $h^{\bar{A}}$ 
being the auxiliary fields. Let us consider an example of such a term when 
$F^{[n]}$ is not zero
only for $n=2$. Then using (\ref{tilfnpin}) in the definition of action
$S$ and solving for the effective action we find:
\begin{eqnarray}
W= F^{[2]} &-& 2 
[(\nabla_{\ibar_1}\nabla_{\jbar_1}f_{\alpha})f^{\alpha}_{\bar p}
G^{i \bar p}D_{i}[(\nabla_{\ibar_2}\nabla_{\jbar_2}f_{\beta})
(\nabla_{\ibar_3}\nabla_{\jbar_3}f^{\beta})]\nonumber\\ &&+
(\nabla_{\ibar_1}\nabla_{\jbar_1}f_{\alpha})f^{\alpha}_{\bar p}
(\nabla_{\ibar_2}\nabla_{\jbar_2}f_{\beta})f^{\beta}_{\bar q}
G^{\imath \bar p}G^{\ell \bar q}R_{\imath{\ibar_3}\ell{\jbar_3}}
\nonumber\\ && + (\nabla_{\ibar_1}\nabla_{\jbar_1}f_{\alpha})
f^{\alpha}(\nabla_{\ibar_2}\nabla_{\jbar_2}f_{\beta})
(\nabla_{\ibar_3}\nabla_{\jbar_3}f^{\beta})]\theta^{\ibar_1}\eta^{\jbar_1}
\theta^{\ibar_2}\eta^{\jbar_2}\theta^{\ibar_3}\eta^{\jbar_3}+\cdots
\label{w3}
\end{eqnarray}
where dots refer to higher order terms and the lower index $f$'s are
$f_1=f^2$ and $f_2=f^1$. $\nabla$'s are reparametrization covariant
derivatives, $D_i$ are K\"ahler covariant derivatives and $G$ and $R$
are respectively the K\"ahler metric and Riemann tensors. This result for
$W$ agrees with the effective field theory calculation presented in
the Appendix up to a closed $3$-form, which is the ambiguity in
the solution of the recursion relation up to this order.

To summarize, in this section we have given a method to construct a solution
to the recursion relations. Moreover we have seen that the recursion 
relations are a consequence of the underlying supersymmetry of the action
defined in (\ref{susys}).
 
\section{Relation with heterotic amplitudes}

In section 3 we anticipated that the genus $g$ string amplitudes
\begin{equation}
A^{g,n}=\langle F_{\mu\nu}^{2}\lambda^{2g-2}\chibar^{2n}\rangle 
\label{agn}
\end{equation}
are related to the topological amplitudes $F^{g}_{\ibar_{1}\cdots\ibar_{n};
\jbar_{1}\cdots\jbar_{n}}$. In this section we are going to
show this by an explicit computation in the general $(2,0)$ case.\footnote{We
will use NSR formalism in the following. It should be also  possible to do this
in the GS formalism by using the methods of  Ref. \cite{berk} where the
topological nature of these amplitudes  should become more transparent.} 

As we explained in 
section 2, the vertex operators for gauginos contain, in the $-1/2$
ghost picture, a combination of spin fields $S_{\alpha}$, 
whose precise form is governed by kinematics. After bosonizing the
space-time fermionic coordinates $\psi^\mu$ by arranging them into
two complex left-moving fermions $\psi_{1,2}=e^{i\phi_{1,2}}$, we have:
\begin{equation}
S_{1} ~=~ {\rm exp}[\frac{i}{2}(\phi_1+\phi_2)]~,
\label{s1}\vspace{1mm}
\end{equation}
\begin{equation}
S_{2}~=~ {\rm exp}[-\frac{i}{2}(\phi_1+\phi_2)]~,
\label{s2}
\end{equation}
Similarly the spin fields $S^{\dot\alpha}$
for the anti-chiral spacetime fermions
$\bar\chi$ 
have the bosonized 
expressions:
\begin{equation}
S^{\dot 1} ~=~ {\rm exp}[\frac{i}{2}(\phi_1-\phi_2)]~,
\label{sdot1}\vspace{1mm}
\end{equation}
\begin{equation}
S^{\dot 2}~=~ {\rm exp}[-\frac{i}{2}(\phi_1-\phi_2)]~,
\label{sdot2}
\end{equation}
As already
mentioned, we also bosonize the $\beta ,\gamma$ system in terms of a free
boson $\varphi$ and the $\eta ,\xi$ system as usual \cite{FMS}.

By ghost charge counting, and recalling that on a
genus $g$ Riemann surface there are also $2g\! -\!2$ additional
insertions of picture-changing operators, 
we have that the total number of insertions
of picture-changing operators is $3g\! -\!3\!+\!n$.

The anti-self-dual part of the gauge field vertex in the $0$ ghost picture
contains the following left-moving fermionic combinations:
$\psi_1\psi_2$, ${\bar\psi}_1{\bar\psi}_2$ and $\psi_1{\bar\psi}_1
+\psi_2{\bar\psi}_2$ depending on the kinematics.  
We will see below following the eq.(\ref{acy1}) that the purely bosonic 
part of the 
gauge field vertices, $\partial X^{\mu}{\bar J}^{a}$, does not contribute
to the amplitude under consideration.

To make the calculation transparent we choose a kinematic
configuration such that $g\!-\!1$ of the $\lambda$'s appear with
$S_1$ and the remaining $g\!-\! 1$ $\lambda$'s with
$S_2$ . Furthermore we take for left-moving part of
one gauge field vertex 
$\psi_1\psi_2$, and for the second one
${\bar\psi}_1{\bar\psi}_2$.
Also, we choose  $n$ of the $\chibar$'s with $S^{\dot 1}$ and
the remaining $n$ with $S^{\dot 2}$.

It will be also convenient to take the $n$ $\bar\chi$'s 
appearing with $S^{\dot 2}$ in the unintegrated form
by inserting $n$ $c$-ghost fields. Correspondingly,
we will have $3g-3+n$ $b$-ghost insertions to agree
with the dimension of the moduli space of genus $g$
Riemann surfaces with $n$ punctures.
Thus we are led to considering the following amplitude:
\begin{eqnarray}
A^{g,n}~&=&~\langle\prod_{k=1}^{3g-3+n}|(\mu_{k}b)|^2
\prod_{i=1}^{g-1} e^{-\frac{1}{2}\varphi}
S_1\Sigma(x_i){\bar J}^{a_i}({\bar x}_i)
\prod_{i=1}^{g-1} e^{-\frac{1}{2}\varphi}
S_2\Sigma(y_i){\bar J}^{b_i}(\bar{y}_i)\nonumber\\ & &
\psi_1\psi_2(z){\bar J}^{a_g}(\bar{z})
{\bar\psi}_1{\bar\psi}_2(w){\bar J}^{b_g}(\bar{w})
\prod_{i=1}^n V^{\dot 1}_{\ibar_i}(u_i,{\bar u}_i)\nonumber\\ && 
\prod_{i=1}^n c{\bar c}V^{\dot 2}_{\jbar_i}(v_i,{\bar v}_i) 
\prod_{i=1}^{3g-3+n} e^{\varphi} T_{F}(z_i){\rangle}_g~,
\label{amplgn}
\end{eqnarray}
\noindent
where the indices $\ibar$ and $\jbar$ label the anti-modulini $\bar\chi$. We
will show now that this amplitude is proportional to the topological amplitude
$F^{g}_{\ibar_{1}\cdots\ibar_{n}; \jbar_{1}\cdots\jbar_{n}}$ for general
$(2,0)$ 
compactifications.

The universal feature of all $(2,0)$ compactifications is the underlying 
$U(1)$ current algebra which can be bosonized in terms
of a free scalar field $H$ \cite{susy}. $H$ is a compact boson and
its momenta sit in a one-dimensional lattice given by the $U(1)$ charges of
the states. Besides the part of the superconformal field theory containing the
space-time fermionic coordinates and the superghosts, the spin structure
dependence enters only through appropriate shifts of the one-dimensional lattice
of $H$. The remaining part of the internal theory does not see the spin
structures. Therefore to do the spin structure sum it is sufficient to know how
it enters in the $U(1)$-charge lattice. On the other hand the topological theory
involves precisely twisting by adding an appropriate background charge for the
field $H$, and again the rest of the internal theory is insensitive to this
twisting. It is this fortunate circumstance that will enable us to show the
equivalence between the string amplitude and $A^{g}_{{\ibar_1}\cdots{\ibar_n}
;{\jbar_1}\cdots{\jbar_n}}$, without ever needing to know the
details of the compactification.

Let $\Gamma$ be the $U(1)$ lattice of $H$ momenta. The space-time fermionic
coordinates define an $SO(2)\times SO(2)$ lattice. If one takes one of these
$SO(2)$ lattices and combines it with $\Gamma$, then it is known that the
resulting $2$-dimensional lattice is given by the coset $E_6/SO(8)$
\cite{Ler,LT}. The characters are given by the branching functions
$F_{\Lambda,s}(\tau)$ satisfying:
\begin{equation}
\chi_{\Lambda}(\tau) = \sum_s F_{\Lambda,s}(\tau)\chi_{s}(\tau)\ ,
\label{chils}
\end{equation}
where $\chi_{\Lambda}$ and $\chi_{s}$ are $E_6$ and $SO(8)$ level one
characters, $\Lambda$ denotes the three classes of $E_6$, and we are
using the spin structure basis denoted by $s$ to represent the four
conjugacy classes of $SO(8)$. The characters of the internal conformal
field theory times one complex space-time fermionic coordinate can then be
represented as $ F_{\Lambda,s}(\tau){\rm Ch}_{\Lambda}(\tau)$, where ${\rm
Ch}_{\Lambda}(\tau)$ is the contribution of the rest of the internal
theory. The essential point here is that ${\rm Ch}_{\Lambda}(\tau)$
depends only on $\Lambda$ and not on the $SO(8)$ representations or
equivalently on the spin structures. Generalization of this to higher
genus is obtained by assigning an $E_6$ representation $\Lambda$ for
each loop and we will denote this collection by $\{ \Lambda\}$.

To proceed further, we note that in the amplitude (\ref{amplgn}), due to the
conservation of $U(1)$ charge only $G^-$ parts
of the $T_F$'s contribute. As before, let us consider the
contribution of the left-moving sector to the amplitude. The dependence of
$G^-$ on $H$ is given by:
\begin{equation}
G^- = e^{-iH/\sqrt{3}} {\hat G}^-\ ,
\label{gminus}
\end{equation}
where ${\hat G}^-$ has no singular operator product expansion with the
$U(1)$ current and it carries a dimension $4/3$. 
Furthermore, the anti-chiral fields $\Psi_{\ibar}$ carry charge
$(-1)$, and therefore their $H$-dependence is given by:
\begin{equation}
\Psi_{\ibar}=e^{-iH/\sqrt{3}} {\hat{\Psi}_{\ibar}},
\label{psi}
\end{equation}
where $\hat{\Psi}$ has non singular OPE with the $U(1)$ current and
has dimension $1/3$. It follows that the internal part 
$\Sigma_{\ibar}$ of the fermion vertex (\ref{fvertex}) is given by:
\begin{equation}
\Sigma_{\ibar}=e^{+iH/2\sqrt{3}} {\hat{\Psi}_{\ibar}}.
\label{sigma}
\end{equation}

Taking the specific kinematic configuration as in (\ref{amplgn}), 
one can now explicitly
compute the correlation functions for the space-time fermions, superghost
and the free field $H$. After combining the lattice of (say) $\psi_2$
with that of $H$, the result can be expressed as:
\begin{eqnarray}
A^{g,n}_{(f)}&=&\frac{\theta_{s}(\frac{1}{2}\sum_{i}(x_i-y_i)+
\frac{1}{2}\sum_{k}(u_k-v_k)+z-w)}
{Z_{1}\theta_{s}(\frac{1}{2}\sum_{i}(x_i+y_i)+\frac{1}{2}\sum_{k}(u_k+v_k)
-\sum_{a}z_a+
2\Delta)} \nonumber\\ & &
F_{\{\Lambda\},s}(\frac{1}{2}\sum_{i}(x_i-y_i)-
\frac{1}{2}\sum_{k}(u_k-v_k) +z-w\ ;\nonumber\\ & & 
\frac{\sqrt 3}{2}\sum_{i}(x_i+y_i)+\frac{1}{2\sqrt 3}\sum_{k}(u_k+v_k)
-\frac{1}{\sqrt 3}
\sum_{a}z_a)\nonumber\\ & &
\frac{\prod_{i<j} E(x_i,x_j) E(y_i,y_j)}  {E^{2}(z,w)\prod_{a<b}
E^{2/3}(z_a,z_b)}
\prod_{i}\frac{E(x_i,z)E(y_i,w)}{E(x_i,w)E(y_i,z)}\nonumber\\ & &
\frac{\prod_{k<\ell}E^{1/3}(u_k,u_{\ell})E^{1/3}(v_k,v_{\ell})}
{\prod_{k,\ell}E^{2/3}(u_k,v_{\ell})}\prod_{k,a}E^{1/3}(u_k,z_a)
E^{1/3}(v_k,z_a)\nonumber\\ &&
\frac{\prod_{i}\sigma(x_i)\sigma(y_i)\prod_k\sigma(u_k)
\sigma(v_k)}{\prod_{a}\sigma^{2}
(z_a)}G_{ \{ \Lambda\} }(\{ z_a,u_k,v_k \} )~, 
\label{acy}
\end{eqnarray}
where $\theta_s$ denotes the genus $g$ $\theta$-function of
spin structure $s$, $E$ is the prime form and $Z_1$ is the
chiral determinant of the $(1,0)$ system. $\Delta$ is the Riemann
$\theta$-constant, which represents the degree $g\!-\!1$ divisor of a half
differential associated with the preferred spin structure for a given marking of
the Riemann surface. $\sigma(x)$ is a $g/2$-differential with no zeros or
poles, transforming in a quasiperiodic way under $a_i$ or $b_i$
monodromies. Finally, the function $G_{ \{ \Lambda\} }$ is given by:
\begin{equation}
G_{ \{ \Lambda\} }(\{ z_a,u_k,v_k \} ) = \langle \prod_a{\hat G}^-(z_a)
\prod_{k}\hat{\Psi}_{\ibar_k}(u_k)\hat{\Psi}_{\jbar_k}(v_k)
\prod_{i} Q^{a_i}_k\bar{\omega}_k(\bar{x}_i)
\prod_{j} Q^{b_j}_k\bar{\omega}_k(\bar{y}_j)\rangle_{ \{ \Lambda\} }\ ,
\label{glambda}
\end{equation}
where $i$ and $j$ run from 1 to $g$, with $x_g=z$ and $y_g=w$, and
$k$ runs over the $g$ anti-holomorphic differentials.
$G_{ \{ \Lambda\} }$ represents the contribution of the internal conformal
field theory after removing the $H$ contribution and does not depend on
spin structure. As explained in section 2 we have replaced the Kac-Moody
currents by their zero-mode parts.
$F_{\{\Lambda\},s}(u;v)$ represents the $SO(2)\times\Gamma$ lattice contribution
to genus g partition function with sources $u$ and $v$; $u$ is coupled to
$SO(2)$ lattice and $v$ is coupled to $\Gamma$ \cite{Ler}.

To do the sum over spin structures we choose the positions $z_a$ such that
\begin{equation}
\sum_a z_a = \sum_i y_i + \sum_k v_k - z + w +2\Delta.
\label{gauge}
\end{equation}
As a result the theta functions in the numerator of eq.(\ref{acy}), cancel with
those in the denominator. The only spin structure dependence then appears in
$F_{\{\Lambda\},s}$. We can sum over the spin structures by using the formula
\cite{Ler}:
\begin{equation}
\sum_s F_{\{ \Lambda\},s}(u;v) = F_{\{ \Lambda\}}(\frac{1}{2}u + 
\frac{\sqrt{3}}{2}v\ ;\ \frac{\sqrt{3}}{2}u - \frac{1}{2}v) \ ,
\label{ssum}
\end{equation}
where
\begin{equation}
F_{\{\Lambda\}}(u;v) = \theta (u) \Theta_{\{\Lambda\}}(v) \ .
\label{flambda}
\end{equation}
$\Theta_{\{\Lambda\}}$ are given by:
\begin{equation}
\Theta_{\{\Lambda\}}(v) = \sum_{n_i \in {\bf Z}} exp\bigl(3 \pi i (n_i +
\frac{\lambda_i}{3})\tau_{ij} (n_j+\frac{\lambda_j}{3}) + 2\pi i\sqrt{3}
(n_i+\frac{\lambda_i}{3})v_i \bigr)\ ,
\label{theta2}
\end{equation}
where $i,j=1\cdots g$, and $\lambda_i = 0,1,2$ depending on $E_6$
conjugacy class $\Lambda_i$ corresponding to ${\bf 1}$, ${\bf{27}}$ and
${\overline{\bf {27}}}$, respectively. They can be expressed as combinations
of level six theta functions of Refs. \cite{Ler,LT}.
Using this formula, (\ref{acy}) becomes:
\begin{eqnarray}
A^{g,n}_{(f)}&=&\frac{\theta(\sum_{i}x_i+z-w)
\Theta_{\{\Lambda\}}(\frac{1}{\sqrt 3}
(\sum_{a}z_a+\sum_k(u_k-2v_k)-3\Delta))}
{Z_{1}}\nonumber\\ & & \frac{\prod_{i<j} E(x_i,x_j)
E(y_i,y_j)}  {E^{2}(z,w)\prod_{a<b} E^{2/3}(z_a,z_b)}
\prod_{i}\frac{E(x_i,z)E(y_i,w)}{E(x_i,w)E(y_i,z)}\nonumber\\ & &
\frac{\prod_{k<\ell}E^{1/3}(u_k,u_{\ell})E^{1/3}(v_k,v_{\ell})}
{\prod_{k,\ell}E^{2/3}(u_k,v_{\ell})}\prod_{k,a}E^{1/3}(u_k,z_a)
E^{1/3}(v_k,z_a)\nonumber\\ &&
\frac{\prod_{i}\sigma(x_i)\sigma(y_i)\prod_k\sigma(u_k)\sigma(v_k)}
{\prod_{a}\sigma^{2} (z_a)}G_{ \{
\Lambda\} }(\{ z_a,u_k,v_k \} )~.
\label{acy1}
\end{eqnarray}

Note that the contribution from the bosonic part of the gauge 
vertices, namely $\partial X^{\mu} \bar{J}^a$, would be the same as 
above except that the argument of the first theta function on the 
right hand side becomes just $\sum_i x_i$ and hence the expression 
vanishes due to Riemann vanishing theorem. 

Using the
bosonization formula for untwisted spin $(1,0)$ and $(2,-1)$ determinants
and using the condition (\ref{gauge}), we see that all the space-time
fermion boson as well as ghost and superghost non-zero mode determinants
cancel. We get finally:
\begin{eqnarray}
A^{g,n}_{(f)}&=&\frac{\prod_{a<b}E^{1/3}(z_a,z_b)\prod_{k<\ell}
E^{4/3}(v_k,v_{\ell})}{\prod_{k,a}E^{2/3}(v_k,z_a)}
\frac{\prod_{k,a}E^{1/3}(u_k,z_a)\prod_{k<\ell}E^{1/3}(u_k,u_{\ell})}
{\prod_{k,\ell}E^{2/3}(u_k,v_{\ell})}\nonumber\\ & &
\frac{\prod_{a}\sigma(z_a)\prod_{k}\sigma(u_k)}
{\prod_{\ell}{\sigma}^{2}(v_{\ell})}
\frac{\Theta_{\{\Lambda\}}(\frac{1}{\sqrt 3}
(\sum_{a}z_a+\sum_k(u_k-2v_k)-3\Delta))}{\det h_a(z_b)}\nonumber\\ &&
G_{\{\Lambda\} }(\{ z_a,u_k,v_k \} ) \frac{\det
\omega_i(x_j,z) \det \omega_i(y_j,w)}{(\det{\rm Im}\tau)^2}\nonumber\\ &&
\mid\det (\mu_a h_b)\mid^2({\makebox{right-moving part}})~.
\label{acy2}
\end{eqnarray}
Here $h_{a}$ denote a basis of $3g-3+n$ quadratic differentials, dual
to $\mu_a$,
which span the cotangent space to $M_{g,n}$. They are allowed to have
simple poles at the punctures $v_k$.
One can check that the expression in (\ref{acy2}) has the correct
conformal properties, namely it transforms as a quadratic differential
in each $z_a$, as a zero form in $v_k$ and as a $1-$form in $u_k$.
As in the case of $F^g$ discussed in section 2, the factor
$(\det {\rm Im}\tau)^{-2}$ is canceled after integrating
over $z$, $w$, $x_i$ and $y_i$.

Consider the factor of (\ref{acy2}) which depends on $z_a$:
\begin{eqnarray}
B(\{ z_a,\} ) &\equiv&\frac{ \prod_{a<b}E^{1/3}(z_a,z_b) 
\prod_{a}\sigma(z_a)
\prod_{k,a}
E^{1/3}(u_k,z_a)}{\prod_{k,a}E^{2/3}(v_k,z_a)}\nonumber\\& &
\Theta_{\{\Lambda\}}(\frac{1}{\sqrt 3}
(\sum_{a}z_a+\sum_k(u_k-2v_k)-3\Delta)) 
G_{ \{\Lambda\} }(\{ z_a,u_k,v_k \} )\ .
\label{bdef}
\end{eqnarray}
It transforms as a quadratic differential in each $z_a$, it is holomorphic with
first order zeroes as $z_a \rightarrow z_b$ for $a\ne b$ (generically),
since $G_{ \{\Lambda\} }(\{ z_a,u_k,v_k \} )$ goes as $(z_a-z_b)^{2/3}$
for $z_a\rightarrow z_b$, and it is totally antisymmetric in $z_a$. 
It has also simple poles as $z_a\rightarrow v_k$, since 
$G_{ \{\Lambda\} }(\{ z_a,u_k,v_k \} )$ goes as $(z_a-v_k)^{-1/3}$ for
$z_a\rightarrow v_k$. This implies that
\begin{equation}
B(\{ z_a \} ) \propto \det h_a(z_b)\ ,
\label{bdet}
\end{equation}
with proportionality constant independent of $z_a$. So, we can 
``bring under the integral" of $\det(\mu_a h_b)$ 
the ratio  $B(\{ z_a \} )/ \det h_a(z_b)$ appearing in (\ref{acy2}).
Integration over $x_i,y_i,z$ and $w$ gives $(g!)\cdot (\det Im\tau)^2
\det Q_k^{a_i}\det Q_k^{b_j}$.
The final result is
\begin{eqnarray}
A^{g,n}_{(f)}&=&\int_{M_{g,n}}\frac{\prod_{k<\ell}
E^{4/3}(v_k,v_{\ell})
\prod_{k<\ell}E^{1/3}(u_k,u_{\ell})\prod_{k}\sigma(u_k)}
{\prod_{k,\ell}E^{2/3}(u_k,v_{\ell})
\prod_{\ell}{\sigma}^{2}(v_{\ell})}\nonumber\\ & &
\int \prod_{a=1}^{3g-3+n}d^2z_a(\prod_{a=1}^{3g-3+n}\mu_a B(\{z\} )
\prod_{a=1}^{3g-3+n}\bar{\mu}_a\bar{b}(\zbar_a))\nonumber\\ &&
\det Q_k^{a_i}\det Q_k^{b_j}({\makebox{right-moving part}})~.
\label{acy3}
\end{eqnarray}
Here in the right moving part we put all the contributions coming
from the bosonic sector which have not been explicitly displayed.

Now let us compare this result with the correlation functions of the
topological theory as given in (\ref{nptf}). 
Once again we extract the $H$ dependence from  $G_-$
and from  the fields $\Psi_{\ibar}$, $\tilde{\Psi}_{\jbar}$, of charges
$-1$ and $2$ respectively,
and explicitly work out its correlation function, remembering that in the
twisted version there is a background charge $\frac{\sqrt 3}{2}\int
R^{(2)}H$ in the action. $3g\! -\! 3\!+\!n$ $G_-$'s appearing in the topological
partition function that are folded with the Beltrami differentials
precisely balance this background charge plus the
net charge of $\Psi$'s and $\tilde{\Psi}$'s, which
is $n$. One easily finds that 
$F^g_{\ibar_1,\dots,\ibar_n;\jbar_1,\dots,\jbar_n} = A^{g,n}/(g!)^2$. 
This completes the proof that the heterotic string amplitudes
under consideration are proportional to the topological
amplitudes of (\ref{nptf}) for general $(2,0)$ compactifications.

\section{Effective Field Theory}

\hspace*{-1mm} 
In this section we discuss relations between the genus $g$ topological 
amplitudes \linebreak
$F^{g}_{\ibar_{1}\cdots\ibar_{n};\jbar_{1}\cdots\jbar_{n}}$
and the physical scattering amplitudes of heterotic superstring theory.
We begin by giving a superfield description of the underlying effective 
lagrangian interactions. These interactions can be understood
as higher-dimensional F-terms. The procedure that we adopt here
is to construct such terms in superconformal supergravity, and to impose 
subsequently the standard $N{=}1$ Poincar\'e gauge-fixing constraints
\cite{kugo}.

In superconformal supergravity, the multiplets are characterized by their
Weyl and chiral weights $(\omega,\nu)$ which specify the properties 
under the dilatations and chiral $U(1)$ transformations. 
Under complex conjugation, $(\omega,\nu)\rightarrow (\omega,-\nu)$.
The chiral superfields containing moduli and matter fields carry weights (0,0). 
One additional chiral superfield, the compensator $\Sigma$
with weights (1,1), is introduced for the sole purpose
of breaking the superconformal group down to $N{=}1$ Poincar\'e supersymmetry,
by constraining its scalar and fermionic components.
An invariant action can be constructed from
the F-component of a chiral superfield with weights (3,3). For example,
the standard superpotential $w$ is a function of (0,0) fields only, and
the corresponding lagrangian density is obtained from the
F-component of $\Sigma^3w$; in this case, the (3,3) weights are supplied
entirely by the chiral compensator. Also, the gauge kinetic terms
can be written as F-terms involving the square
of the canonical gauge field-strength $W$ carrying weights $(3/2,3/2)$; 
the (3,3) weights are then supplied by $W^2$.

Higher-dimensional F-terms can be constructed by including superfields 
which are chiral projections of complex vector superfields. 
The superconformal chiral projection $\Pi$, which is a generalization of the 
$\bar{D}^2$ operator of rigid supersymmetry,
can be defined for vector superfields $V$ with weights (2,0) only;
a chiral superfield $\Pi (V)$ has weights (3,3)
\cite{kugo}. It is convenient to define \cite{mu}
\begin{equation}
\Pi_{\alpha}~\equiv~\Pi(\Sigma\overline{\Sigma}\,e^{-K/3}f^{\alpha})\, ,
\label{pi}\end{equation}
where $K$ is the tree-level K\"ahler potential and
$f^{\alpha}$ are arbitrary functions of chiral and/or
anti-chiral (0,0) superfields. The factor $e^{-K/3}$ has been introduced
to ensure covariant transformation properties of $\Pi$'s under
the K\"ahler transformations $K\rightarrow K+\ph+\bar{\ph}$,
$\Sigma\rightarrow e^{\ph /3}\Sigma$; $\Pi$'s transform then
with the same holomorphic weights as the defining $f$'s.
Let us introduce a class of chiral $(3,3)$ superfields
\begin{equation}
{\cal I}^{g}_{n}~=~ F^g_n\cdot\Sigma^{3(1-g-n)}
\cdot\Pi_1\Pi_2\cdots\Pi_n\cdot
W^{2g}\ , \label{ig}
\end{equation}
where $F^g_n$ are arbitrary {\em analytic} functions of $(0,0)$ chiral 
superfields. We will see that the topological amplitudes are related to 
the interactions obtained from F-terms of such superfields.
Note that Peccei-Quinn symmetry of the effective action 
forbids the dilaton-dependence of $F^g_n$, with the exception of 
tree-level gauge kinetic terms contained in $F^1_0$.

In order to determine the string loop order 
of ${\cal I}^{g}_{n}$, the powers of the string coupling
constant should be counted in the following way. 
In the superconformal theory, the Planck mass squared $M_P^2$ corresponds
to the vacuum expectation value of the scalar component of
$\Sigma\overline{\Sigma}e^{-K/3}$. 
The gauge fixing constraint on the scalar component of $\Sigma$
that results in a sigma-model type normalization
$M_P^2\sim 1/g_s^2$, with $g_s$ the four-dimensional string coupling,
corresponds to $\Sigma\sim e^{K/6}g_s^{-1}$.{}From the known 
dilaton-dependence
of $K$ it follows that
$K\sim\ln g_s^2$, therefore $\Sigma\sim g_s^{-2/3}$. Assuming that
the functions $F^g_n$ and $f^{\alpha}$ do not depend on the dilaton, we
have $\Pi_{\alpha}\sim g_s^{-2}$ and ${\cal I}^{g}_{n}\sim g_s^{2g-2}$.
Hence ${\cal I}^{g}_{n}$ describe $g$-loop string interactions.

To obtain explicit expressions for particle interactions, we will write down
the component content of superfields $\Sigma$ and $\Pi$.
In general, a chiral superfield $\Phi=(z_{\Phi},{\cal Z}_{\Phi},
{\cal F}_{\Phi})$, where $z$, $\cal Z$ and $\cal F$ are the
scalar, fermionic and auxiliary components, respectively.
The multiplication rule is 
\begin{equation}
(z_1,{\cal Z}_1,{\cal F}_1)\times (z_2,{\cal Z}_2,{\cal F}_2)=
(z_1z_2,\, z_1{\cal Z}_2+ z_2{\cal Z}_1,\, z_1{\cal F}_2+z_2{\cal F}_1-
2{\cal Z}_1{\cal Z}_2).    \label{multi}
\end{equation}
After superconformal gauge fixing,\footnote{From now on
we set the string coupling $g_s=1$.} the compensator becomes
$\Sigma=e^{K/6}(1,{\cal Z}_{\Sigma},{\cal F}_{\Sigma})$, with
\begin{eqnarray}
{\cal Z}_{\Sigma} &=& {1\over 3}K_i\chi^i\label{zsigma}\\
{\cal F}_{\Sigma} &=& h^0+{1\over 3}K_i
(h^i+\Gamma^i_{jk}\chi^j\chi^k)-{1\over 3}(K_{ij}+
{1\over 3}K_iK_j)\chi^i\chi^j\ . \label{fsigma}
\end{eqnarray}
Here, $\chi^i$ and $h^i$ denote the fermionic and auxiliary components,
respectively, of the physical moduli and matter fields. $h^0$ is an auxiliary
field left over from the compensator; it is usually
eliminated from the effective lagrangian together with all other
auxiliary fields by using their equations of motion. The reparametrization 
connection $\Gamma^k_{ij}\equiv K^{-1\,k\lbar}K_{ij\lbar}$. 

The components of $\Pi(\Sigma\overline{\Sigma}\,e^{-K/3}f)$ are \cite{mu}:
\begin{eqnarray}
z_{\Pi}&=& \chibar^{\ibar}\chibar^{\jbar}\nabla_{\ibar}f_{\jbar}
-h^{\bar{0}}f-h^{\ibar}f_{\ibar}\label{zpi}\\
{\cal Z}_{\Pi}&=&f_{\ibar\jbar}\dslash z^{\ibar}\chibar^{\jbar}+
f_{\ibar}\dslash\chibar^{\ibar}\nonumber\\ & &+
{f\over 3} (K_{i\jbar}\dslash z^{i}\chibar^{\jbar}-2\sigma^{mn}\partial_m\psi_n)
h^{\bar{0}}f_i\chi^i
-{1\over 3}(K_{\ibar j}f-3f_{\ibar j})h^{\ibar}\chi^j+\dots\label{ferpi}\\
{\cal F}_{\Pi}&=& -f_{\ibar\jbar}\partial_m z^{\ibar}
\partial_m z^{\jbar}
-f_{\ibar}\partial^2 z^{\ibar}-{f\over 6}(2K_{i\jbar}\partial_m z^{i}
\partial_m z^{\jbar}+R)\nonumber\\ & &
-h^0h^{\bar{0}}f+{1\over 3}(K_{\ibar j}f-3f_{\ibar j})h^{\ibar}h^j
-f_{\ibar}h^{\ibar}h^0 - f_{i}h^{i}h^{\bar{0}}\nonumber\\ & &
+  \{h^0\nabla_{\ibar}f_{\jbar}\chibar^{\ibar}
\chibar^{\jbar}+h^k\nabla_{\ibar}f_{\jbar k}\chibar^{\ibar}
\chibar^{\jbar}-{2\over 3}h^k\chibar^{\ibar}\chibar^{\jbar}
K_{k\ibar}f_{\jbar}+\rm{c.c.}\}+\dots\label{fpi}
\end{eqnarray}
where $\psi_m$ is the gravitino field.
Here, the reparametrization covariant derivative $\nabla$ acts as
$\nabla_{\ibar}f_{\jbar}=f_{\ibar\jbar}-\Gamma^{\kbar}_{\ibar\jbar}
f_{\kbar}$. The terms neglected in eqs.(\ref{ferpi}) and (\ref{fpi})
are not relevant to the following discussion. 
We complete our list by recalling the component content
of the gauge kinetic multiplet:
\begin{equation}
W^2=[\;\lambda\lambda,~ F_{mn}\sigma^{mn}\lambda+\dots,~ 
\frac{1}{2}(F_{mn}F^{mn}+iF_{mn}\tilde{F}^{mn})+\dots\;]\ ,  \label{wcomp}
\end{equation}
where $\lambda$ and $F_{mn}$ are the gaugino and gauge field strength,
respectively. Here again, we omitted some irrelevant terms.

The F-term lagrangian density obtained from a (3,3) chiral superfield $\Phi$
is given by the standard formula \cite{kugo}:
\begin{equation}
\Phi|_F={\cal F}_{\Phi}+\bar{\psi}_m\bar{\sigma}^m{\cal Z}_{\Phi}
+\bar{\psi}_m\bar{\sigma}^{mn}\bar{\psi}_n\, z_{\Phi} +\rm c.c.\label{fphi}
\end{equation}
Note that the scalar component
of $\Phi$ gives rise to a field-dependent gravitino mass.
By using the expressions (\ref{zsigma})-(\ref{wcomp}), 
the multiplication rule (\ref{multi}) and the above formula,
one can derive explicit form of the interactions induced
by ${\cal I}^g_n$ of eq.(\ref{ig}). It is worth mentioning that the tree-level
kinetic energy terms, which are usually written as D-terms, can also
be written as the F-term of 
\begin{equation}
{\cal I}^0_1=   3\,\Pi(\Sigma\overline{\Sigma}\,e^{-K/3})\ ,\label{i0}
\end{equation}
{\em c.f}.\ eqs.(\ref{fphi}) and (\ref{fpi}). 

The genus $g=0$ topological amplitudes
are related to the tree-level interactions ${\cal I}^0_n$, $n>1$,
in the following way. Let us consider the tree-level scattering amplitude
\begin{equation}
\lv\chibar^{\ibar_1},\chibar^{\ibar_2},\dots,
\chibar^{\ibar_{n-1}};\chibar^{\jbar_1},\chibar^{\jbar_2},\dots,
\chibar^{\jbar_{n-1}};z^{\ibar_n},z^{\jbar_n}\rule{0mm}{4mm}\rv
~=~ p^{\ibar_n}\!\cdot \! p^{\jbar_n}\, A^{0,n}\ ,\label{mamp}
\end{equation}
where $A^{0,n}$ is a momentum-independent function of
the background moduli fields. 
The helicity configuration is chosen in such a way that
$(\chibar^{\ibar_{\alpha}}_{\dot{1}},\chibar^{\ibar_{\alpha}}_{\dot{2}})
= (\theta^{\ibar_{\alpha}},0)$ and
$(\chibar^{\jbar_{\alpha}}_{\dot{1}},\chibar^{\jbar_{\alpha}}_{\dot{2}})
= (0,\eta^{\jbar_{\alpha}})$. This amplitude can be generated
by the interaction terms contained in ${\cal I}^0_n$
and ${\cal I}^0_m,~m<n$. 
The effective interaction term that we are interested in has the form
\begin{equation}
e^{\frac{(1-n)}{2}K}F^{0}_{\ibar_1\ibar_2\dots\ibar_{n};
\jbar_1\jbar_2\dots\jbar_{n}}
\theta^{\ibar_1}\theta^{\ibar_2}\dots\theta^{\ibar_{n-1}}
\eta^{\jbar_1}\eta^{\jbar_2}\dots\eta^{\jbar_{n-1}}\,
\partial_p {\zbar}^{\ibar_n}\partial^p {\zbar}^{\jbar_n}\ . \label{finn}
\end{equation}
The functions $F^{0}_{\ibar_1\ibar_2\dots\ibar_{n};
\jbar_1\jbar_2\dots\jbar_{n}}$ are given by the tree-level topological
amplitudes considered in the previous section.

Let us first consider
\begin{equation}
{\cal I}^{0}_{n}|_F~=~F^0_n\cdot\Sigma^{3(1-n)}\cdot\Pi_1\Pi_2\cdots\Pi_n|_F\ .
\label{iof}
\end{equation}
${\cal I}^{0}_{n}|_F$ gives rise to two basic
contributions to eq.(\ref{finn}). The first one is due to the terms of the form
$z_{\Pi_1}z_{\Pi_2}\dots{\cal F}_{\Pi_k}\dots z_{\Pi_n}$, with ${\cal F}_{\Pi}$
and $z_{\Pi}$'s contributing $\partial \zbar\partial\zbar$ 
and fermion bilinears, respectively, see eqs.(\ref{zpi}) and (\ref{fpi}).
The second one is due to the terms of the form
$z_{\Pi_1}z_{\Pi_2}\dots{\cal Z}_{\Pi_k}{\cal Z}_{\Pi_l}\dots z_{\Pi_n}$, with
both ${\cal Z}_{\Pi}$'s contributing $\dslash\zbar\chibar$, 
see eq.(\ref{ferpi}), and $z_{\Pi}$'s as before. After Fierz transformation, 
this contribution acquires the desired form of eq.(\ref{finn}).
The physical amplitudes receive also additional contributions
due to reducible diagrams with
scalar and/or fermion propagators attached to the operators
$\partial^2\zbar$ and $\displaystyle{\dslash}\chibar$ contained in 
${\cal F}_{\Pi}$'s and ${\cal Z}_{\Pi}$'s;
after including these diagrams, the effective action becomes
manifestly field-reparametrization invariant. Combining everything
together yields
\begin{equation}
{}F^{0}_{\ibar_1\ibar_2\dots\ibar_{n};\jbar_1\jbar_2\dots\jbar_{n}}
=F^{0}_n\,\sum_{\sigma,\omega}\,(-1)^{{\rm sgn}(\sigma )+{\rm sgn}(\omega )}
\,f^{1}_{\ibar_{\sigma(1)}\jbar_{\omega(1)}}
f^{2}_{\ibar_{\sigma(2)}\jbar_{\omega(2)}}\dots
f^{n}_{\ibar_{\sigma(n)}\jbar_{\omega(n)}}\ , \label{pino}
\end{equation}
where the summation extends over all permutations $\sigma$ and $\omega$
of $n$ indices $\ibar$ and $\jbar$, respectively. 
As a result, $F^0$ is completely antisymmetric
in $\ibar$ and $\jbar$ indices separately. It also satisfies trivially the
identity (\ref{ident}). Furthermore, it satisfies
\begin{equation}
\nabla_{[\ibar}F^{0}_{\ibar_1\ibar_2\dots\ibar_{n}];
\jbar_1\jbar_2\dots\jbar_{n}}=0\ , \label{equ}
\end{equation}
due to K\"ahler geometry and analyticity of $F^{0}_n$, as required
by supersymmetry.

Before explaining how ${\cal I}^0_m$, with $m<n$,
affect $2n$-point topological amplitudes
we would like to point out another class of related scattering amplitudes.
As already mentioned before, see eq.(\ref{fphi}), locally
supersymmetric F-terms give rise to field-dependent gravitino
mass terms. These contribute to the amplitude
\begin{equation}
\lv\chibar^{\ibar_1},\chibar^{\ibar_2},\dots,
\chibar^{\ibar_{n}};\chibar^{\jbar_1},\chibar^{\jbar_2},\dots,
\chibar^{\jbar_{n}};\bar{\psi}_{p\,\dot{\alpha}},\bar{\psi}_{q}^{\dot{\beta}}
\rule{0mm}{4mm}\rv
~=~ \bar{\sigma}^{pq\,\dot{\alpha}}_{~~~\dot{\beta}}\:
M^{0,n}\ ,\label{psiamp}
\end{equation}
where $M^{0,n}$ is a momentum-independent function of the background
moduli fields. Here again, the helicity configuration
is chosen in the same way as in the amplitude (\ref{mamp}).
The corresponding effective action term has the form
\begin{equation}
{\textstyle\frac{1}{(n!)^2}}
e^{\frac{(1-n)}{2}K}F^{0}_{\ibar_1\ibar_2\dots\ibar_{n};
\jbar_1\jbar_2\dots\jbar_{n}}
\theta^{\ibar_1}\theta^{\ibar_2}\dots\theta^{\ibar_{n}}
\eta^{\jbar_1}\eta^{\jbar_2}\dots\eta^{\jbar_{n}}\:
\bar{\psi}_p\bar{\sigma}^{pq}\bar{\psi}_q\ .
\label{psinn}
\end{equation}
The fact that the same (topological) functions $F^0$ appear above as in the
scalar-fermion interactions of eq.(\ref{finn}) is a direct consequence
of supersymmetry. It is easy to see that ${\cal I}^0_n|_F$ gives
rise to an effective gravitino mass of the form (\ref{psinn}),
with the coefficient given by eq.(\ref{pino}).

The scattering amplitudes of eqs.(\ref{mamp}) and (\ref{psiamp}),
and hence the respective $2n$-point topological amplitudes,
receive also contributions from the interactions induced by
${\cal I}^0_m$ with $m<n$. In general, these interactions
arise from reducible diagrams with the auxiliary $h$-fields 
propagating on internal lines; these may combine with
diagrams that are reducible in the physical scalar and fermion lines.
Let us illustrate this point by constructing a typical
contribution to the fermion-scalar amplitude (\ref{mamp}).
${\cal I}^0_m|_F$  induces an interaction term of the form
$\partial_iF^{0}_m\, h^i(\chibar\chibar)^m$. The auxiliary field
$h^i$ can propagate to the vertex $h^{\ibar}(\partial \zbar\partial\zbar)
(\chibar\chibar)^{m'-2}$ induced by (possibly the same) ${\cal I}^0_{m'}|_F$,
producing an effective $(\chibar\chibar)^{m+m'-2}(\partial
\zbar\partial\zbar)$ coupling which has the same structure
as the one induced directly by ${\cal I}^0_n|_F$ with
$n=m+m'-1$. This is only one of quite a few reducible diagrams.
Note also that in general, due to the presence of
higher power auxiliary field interactions, $h\bar{h}^m$ ($m>1$), 
their elimination can be highly non-trivial.
In Appendix A, we show explicitly how to evaluate ${\cal I}^0_3\sim\Pi^3$-type
terms induced by ${\cal I}^0_2|_F\sim\Pi^2$, however it is not
clear how to proceed in more complicated cases.

To recapitulate, the computation of the effects of higher-dimensional
F-terms on the scattering 
amplitudes involves elimination of auxiliary fields. 
As a result, the complete functions 
${}F^{0}_{\ibar_1\ibar_2\dots\ibar_{n};\jbar_1\jbar_2\dots\jbar_{n}}$
no longer satisfy eq.(\ref{equ}). Instead, they satisfy
the recursion relations (\ref{hanomn}) (in which
$g=0$) obtained in section 3 by
using topological methods which capture also,
in a simple way, the effects of complicated field-theoretical diagrams.
These relations are not sufficient though to determine all
functions since the solutions can only be determined
up to holomorphic pieces that satisfy the homogeneous
eq.(\ref{equ}) as discussed in section 4. 
Thus the analytic coefficients $F^0_n$ of $\Pi^n$ terms remain arbitrary,
whereas non-holomorphic parts are constrained by eqs.(\ref{hanomn}).
We are not able to prove here rigorously
that the elimination of auxiliary fields in $\Pi^n$-type terms
yields precisely the same equations.
The example discussed in Appendix A provides some circumstantial evidence
that this is indeed the case.
Further support is provided by the interpretation of the tree-level recursion
relation as a supersymmetric Ward identity (\ref{ward}). 

In section 4, we presented an algorithm to generate non trivial solutions
of the recursion relations. It employs a Ward identity associated to the
transformations (\ref{susys}). From the supergravity point of view, these
transformations correspond to the residual supersymmetry in a special type of
superconformal gauge, $\Sigma =(1,0,h^0)$. In this gauge, the standard
supersymmetry transformation $\delta_Q(\epsilon)$ \cite{kugo}
must be followed by a compensating $S$ supersymmetry transformation
$\delta_S(\zeta=\epsilon h^0)$ which is necessary to restore the gauge.
Such a combined transformation which preserves the form
of the superconformal gauge fixing condition defines 
Poincar\'e supersymmetry. The transformation
$\delta_s(\varepsilon)$ (\ref{susys}) corresponds to ``1/2" of Poincar\'e
supersymmetry, namely with $\epsilon=0$,
$\bar{\epsilon}=(0,\bar{\varepsilon})$ and with all spacetime derivative terms
set to zero. It is not surprising that this transformation is not nilpotent on
the fields, since the anticommutator of $Q$ and $S$ generators produces
additional dilatations and axial rotations.

Although from the topological point of view, eq.(\ref{hanomn})
describes an anomaly, there is clearly no field-theoretical
anomaly at the tree-level.  Such an anomaly, the so-called holomorphic anomaly,
appears though in one-loop threshold corrections to gauge couplings.
It shows up in higher genus recursion relations, as explained below.

The higher genus F-terms, ${\cal I}^g_n\sim F^{g}_n\Pi^nW^{2g}$, 
can be discussed in a similar
way. In sections 2 and 3 we have considered genus $g$ amplitudes
\begin{eqnarray}
\lv\chibar^{\ibar_1},\dots,\chibar^{\ibar_{n}};\chibar^{\jbar_1},\dots,
\chibar^{\jbar_{n}};\lambda^{a_1}_{\alpha_1},\lambda^{b_1}_{\beta_1},\dots,
\lambda^{a_{g-1}}_{\alpha_{g-1}},\lambda^{b_{g-1}}_{\beta_{g-1}};
F^{a_g}_{mn},F^{b_g}_{pq}\rule{0mm}{4mm}\rv & &\nonumber\\ & &\hspace{-6cm}
=~ A^{g,n}\:\delta_{mp}\delta_{nq}\,\delta^{a_gb_g}
\prod_{k=1}^{g-1}\delta^{a_kb_k}\epsilon_{\alpha^k\beta^k}
\end{eqnarray}
with the usual helicity configuration of $\chibar$'s.
The corresponding effective action term,
\begin{equation}
e^{\frac{(1-g-n)}{2}K}gF^{g}_{\ibar_1\ibar_2\dots\ibar_{n};
\jbar_1\jbar_2\dots\jbar_{n}}
\theta^{\ibar_1}\theta^{\ibar_2}\dots\theta^{\ibar_{n}}
\eta^{\jbar_1}\eta^{\jbar_2}\dots\eta^{\jbar_{n}}\:
(\lambda\lambda)^{g-1}\:F_{mn}F^{mn}\ , \label{fgi}
\end{equation}
can be induced either directly by ${\cal I}^{g}_n$ or via reducible diagrams 
involving couplings induced by ${\cal I}^{g'}_{n'}$ with $n'<n$ and/or $g'<g$.
The one-loop threshold correction $F^1$ receives also contributions
from the anomalous, non-local effective action terms that violate
holomorphicity of the field-dependent gauge couplings.
These anomalies can feed into higher genus through reducible diagrams, 
as illustrated in Appendix B, where a one-loop ${\cal I}^1_1\sim\Pi W^2$-type
term is induced by a tree-level ${\cal I}^0_2|_F\sim\Pi^2$ interaction, via
one-loop, anomalous $W^2$ couplings. Also in the next section we show by an
explicit string computation on a simple orbifold example, that a tree-level
$\Pi^3$ interaction generates a sequence of non-holomorphic terms: $\Pi^2 W^2$
at the one-loop, $\Pi W^4$ at two loops, and $W^6$ at three loops.

We should finally mention that as in the tree-level case, 
the F-term interactions (\ref{fgi}) have locally supersymmetric completions 
which include field-dependent gravitino mass terms,
\begin{equation}
e^{\frac{(1-g-n)}{2}K}F^{g}_{\ibar_1\ibar_2\dots\ibar_{n};
\jbar_1\jbar_2\dots\jbar_{n}}
\theta^{\ibar_1}\theta^{\ibar_2}\dots\theta^{\ibar_{n}}
\eta^{\jbar_1}\eta^{\jbar_2}\dots\eta^{\jbar_{n}}\:
(\lambda\lambda)^{g}\:\bar{\psi}_p\bar{\sigma}^{pq}\bar{\psi}_q\ .\label{fgij}
\end{equation}

As discussed in Sections 2 and 3, from the topological
point of view, the recursion relations (\ref{hanomn})
reflect anomalies of the underlying two-dimensional, twisted superconformal
field theory. These anomalies are also reflected in the scattering
amplitudes of massless superstring excitations in four dimensions.
Here, within the framework of effective field theory,
these relations describe an intricate interplay within a large class
of higher-dimensional F-terms (\ref{ig}) and can be understood as a consequence
of $N=1$ Poicar\'e supersymmetry.

\section{Orbifold examples}

Here we will work out some simple examples in orbifold models 
of the topological quantities   
we have introduced in sections 2 and 3.   

In the case of orbifolds, the internal $N=2$ SCFT is realized
in terms of free bosons and fermions. We consider for simplicity
orbifolds realized in terms of $3$ complex bosons $X_{I}$ and
left-moving fermions $\Psi_{I}$, 
with $I=1,2,3$, together with $16$ right-moving bosons
living on the $E_{8}\times E_{8}$ lattice. 
Let $h$ be an element of the orbifold group
defined by $h=\{h_{I},\delta_{h}\}$, and its action on $X_I$ is
$X_{I}\rightarrow e^{2\pi i h_{I}}X_I$ and similarly for
$\Psi_I$ and $\delta_{h}$ denotes a shift on the
$16$ right-moving bosons. Space-time supersymmetry implies that
one can always choose the $h_I$'s to satisfy the condition:
\begin{equation}
\sum_{I} h_{I}~=~0\ .
\label{susycond}
\end{equation}
On a genus $g$ Riemann surface we must associate
to each homology cycle $a_{i},b_{i}$, for $i=1,\cdots,g$, an element
of the orbifold group. In the following we shall denote by
$\{h\}=\{\{h_I,\delta_{h}\}\}$ the set of all twists along different cycles.
One can bosonize the complex fermions
\begin{equation}
\Psi_{I}=e^{i \Phi_I}~,~~~~~~~~~~~~~\psibar_{I}=e^{-i\Phi_I}~,
\label{boso}
\end{equation}
The previously introduced free boson $H$ (see section 5) can be expressed
in terms of the $\Phi_{I}$'s as:
$\sqrt{3}H=\sum_{I=1}^{3}{\Phi}_{I}$. We also need the form of vertex
operators for the untwisted anti-moduli ${\tbar}_I$:
\begin{equation}
V_{\tbar_I}=(\partial {\xbar_I}+ip\cdot\psi\psibar_I)
\bar{\partial}{ X}_Ie^{ip\cdot X}\ .
\label{vit}
\end{equation}
In the case of a ${\bf Z}_2$ twisted plane, there is also an 
associated $U$-modulus. The ${\ubar}_I$ vertex operator  $V_{\ubar_I}$
is obtained
from the above expression by replacing $\bar{\partial}{ X}^{I}$ with
$\bar{\partial}{\xbar}^{I}$. Similarly, the vertex for an 
untwisted $\chibar_{\tbar_I}$ 
is given by:
\begin{equation}
V^{\dot\alpha}_{\tbar_I}=(e^{-\frac{\varphi}{2}}S^{\dot\alpha}
e^{-\frac{\Phi_{I}}{2}+\frac{1}{2}\sum_{J\ne I}\Phi_{J}})(z)
\bar{\partial}{ X}^{I}(\bar{z})e^{ip\cdot X}\ ,
\label{vichi}
\end{equation}
with a similar expression for $V^{\dot\alpha}_{\ubar_I}$.

At the tree-level, the simplest topological
quantity is of the type $\Pi^2$ which was considered in Ref. \cite{mu}. However,
it was shown there to be zero for untwisted moduli whereas for twisted
moduli it was shown to be non-vanishing in an explicit example of 
${\bf Z}_6$ orbifold in Ref. \cite{pi2}. For untwisted moduli, the simplest
non-vanishing topological quantity is of the type $\Pi^3$. This
corresponds to a 6-point amplitude (\ref{mamp}) involving two
anti-moduli (\ref{vit}) and four anti-modulini (\ref{vichi}). 
Using the explicit form of the vertices and left-moving
charge conservation, one can easily see that a non-vanishing
contribution is obtained only for ${\bf Z}_2$ twists and when 
both $T$- and $U$-type fields are present for all three planes.
Moreover, the vertices of the two anti-moduli contribute only with their
fermionic part providing the two space-time momenta. In an appropriate
kinematic configuration,
the result can be recast in the form of the topological amplitude:
\begin{eqnarray}
& &F^0_{\tbar_1\tbar_2\tbar_3;\ubar_1\ubar_2\ubar_3}= \nonumber\\
& & \langle \dbar X_1\psibar_1(z_1)
\dbar X_2\psibar_2(z_2) \dbar X_3\psibar_3(z_3)
\dbar\xbar_1\Psi_2\Psi_3(w_1)
\dbar\xbar_2\Psi_3\Psi_1(w_2)
\dbar\xbar_3\Psi_1\Psi_2(w_3)  \rangle\ ,
\label{F0123top}
\end{eqnarray}
where $\Psi_I$ and $\psibar_I$ ($I=1,2,3$) are, respectively, 
zero- and one-forms in the topological theory.
Note that the fermion bilinears $\Psi_I\Psi_J$ appearing in the expression
(\ref{F0123top}) are precisely the left moving parts of the
(left) charge 2 and dimension (0,0) operators ${\tilde\Psi}_{\bar\jmath}$
defined in eq.(\ref{nptf}) with the holomorphic three-form
$\rho=\Psi_1\Psi_2\Psi_3$. The final result is:
\begin{equation}
F^0_{\tbar_1\tbar_2\tbar_3;\ubar_1\ubar_2\ubar_3}=
\prod_{I=1}^3{1\over (T_I+\tbar_I)^2(U_I+\ubar_I)^2}\ ,
\label{f0123}
\end{equation}
while all other non-vanishing amplitudes can be obtained by using
the antisymmetry properties of $F^0$ and the symmetry under
interchanging $T$- and $U$-moduli of the same plane. Notice that this
expression is covariantly constant with respect to all 6 moduli consistently
with the absence of a $\Pi^2$ term.

Eq.(\ref{f0123}) 
can be integrated by using eq.(\ref{pino}) to obtain the functions $f^{1,2,3}$
which enter into the $\Pi_1\Pi_2\Pi_3$ interaction term.
However, this integration is not unique \cite{mu}
because all physical amplitudes involve at least two anti-analytic
derivatives of each $f$. A consistent solution is that $F^0_3=1$ and that $f^I$
depends only on the $T$ and $U$ moduli of the $I$-th plane with
\begin{equation}
f^I_{\tbar_I\ubar_I}={1\over (T_I+\tbar_I)^2(U_I+\ubar_I)^2}\ .
\label{fI}
\end{equation}
One can now use the $SL(2,Z)^6$ duality symmetry associated with
the 6 moduli to integrate eq.(\ref{fI}). $SL(2,Z)$ symmetries induce K\"ahler
transformations which can be compensated by transforming $f$'s.
It is easy to see that $\Pi_I$ must transform as a form of weight 2 with
respect to the $SL(2,Z)$'s corresponding to $T_I$ and $U_I$. This implies
that 
\begin{equation}
f^I=G_2(T_I)G_2(U_I)
\end{equation}
where $G_2(T)\equiv 1/(T+\tbar )+2\partial_T\ln\eta(iT)$. Note that the
analytic part of $G_2$ drops from all physical tree-level amplitudes
consistently with the Peccei-Quinn symmetries of the untwisted moduli.

To discuss examples of topological terms of the type $W^{2g}\Pi^n$ we will
consider models having at least two different pure gauge groups with no massless
matter representations, so that one does not get contributions to the recursion
relations from handle degeneration. A simple way to obtain such models using
fermionic constructions of ${\bf Z}_2\times{\bf Z}_2$ orbifolds 
is by introducing more than one sets of periodic
right-moving fermions \cite{ferm}. 
Each set should contain a number of real fermions
which must be multiple of 8 due to world-sheet modular invariance. In this way,
with two such sets, one gets from the Neveu-Schwarz sector gauge bosons 
in the adjoint representation of $O(8k)\times O(8l)$ (with $k+l\le 5$)
and no massless states in vector representations. 
For the simplest choice, $k=3$ and $l=2$, we obtain $O(24)\times E_8$ with no 
massless matter. However in this case it is easy to see that there are only two
moduli associated to one of the 3 internal planes. To construct families of
models
with at least 3 untwisted moduli associated with 3 different planes one should 
choose for instance $k=3$ and $l=1$.
In this case, a potential problem is the generic appearance of massless
$O(8)$ spinors coming from the corresponding Ramond sector.
However, we checked in several examples that these states can be eliminated 
by an appropriate choice of GSO projections among the various sets of the basis
vectors, leading to a pure gauge group $O(24)\times O(8)$.
          
Going back to the 6 moduli example we considered before, one sees from the
structure of the recursion relations (\ref{hanomn}) that a $\Pi^3$ term on the
sphere generates a $W^2\Pi^2$ term on the torus, a $W^4\Pi$ term on genus 2
and a $W^6$ term on genus 3 with non holomorphic couplings. In the following,
we will present a string derivation of these terms and check the form of the
recursion relations.

Let us start from the $W^2\Pi^2$ term at genus 1. We can, for
instance, take the highest component of the $\Pi^2$ superfield
times the lowest component of $W^2$: this will lead us to consider
an amplitude involving two gauginos, two anti-moduli and
two anti-modulini.  It is easy to see, by left-moving charge
conservation, that the two anti-moduli  and two anti-modulini   
should come from two
different planes.  We will take the anti-moduli to be $\tbar_1$ and $\ubar_2$
and the  anti-modulini  to be the supersymmetric 
partners of $\ubar_1$ and $\tbar_2$. Choosing a convenient kinematic
 configuration, we are thus lead to evaluate the string amplitude:
\begin{equation}
A^{1,2}=\langle V^1_{\lambda}(x) V^2_{\lambda}(y) V^{\dot 1}_{\ubar_1}(z)
V^{\dot 2}_{\tbar_2}(w) V_{\tbar_1}(u) V_{\ubar_2}(v) e^{\varphi}T_F(z_1)
e^{\varphi}T_F(z_2)\rangle\ ,
\label{a12}
\end{equation}
where $V^\alpha_{\lambda}$ is the gaugino vertex operator (\ref{vgn}).
Here we assume that the third plane is untwisted. Using internal left-moving
charge conservation one can show that only the fermionic part of the
anti-moduli vertices
contribute to the amplitude providing also the two factors of
space-time momenta. One can then set the momenta to zero in the
remaining part of the vertices. The spin structures 
sum can be  performed as
usual by using the Riemann theta-identity.  The resulting
expression can be written as a topological amplitude:
\begin{equation}
F^1_{\tbar1,\tbar2;\ubar1,\ubar2}=\langle \dbar X_1\psibar_1(z)
\dbar X_2\psibar_2(v) \dbar\xbar_2\Psi_3\Psi_1(w)
\dbar\xbar_1\Psi_2\Psi_3(u) G^-(z_1)G^-(z_2)\delta Q^2 \rangle\ ,
\label{F112}
\end{equation}
where $\delta Q^2$ denotes the difference between the squared-charges
of the two gauge groups.
$G^-$ is the dimension $(2,0)$ current $G^-=\sum_I\psibar_I\partial X_I$
and by charge conservation we see that only the part corresponding to
the third plane contributes to the amplitude we are considering.  
Furthermore since this plane is 
untwisted, $\partial X_3$ is replaced by its zero-mode $p_L$. 

One can now perform the integrations over the positions of the vertices,
using the theta-functions expression 
for the correlator of the scalars $X_{1,2}$
twisted by $g_{1,2}=g$,
$\langle\dbar \xbar_{1,2} \dbar X_{1,2}\rangle$,  
and doing partial integrations in  the $\bar z$ and $\bar v$
variables.   One is then  left with a final integral of the square of
the Szego kernel corresponding to the twist $g$, which
can again be performed  by a partial integration, giving
a factor $\tau_2\partial _{\bar\tau}\ln(\tau_2\bar{\theta}_g)$,
where $\theta_g$ is the theta-function with characteristics given
by the twist $g$. 
The final expression for the amplitude is reduced to an integral over
the moduli space of the torus:
\begin{eqnarray}
F^1_{\tbar_1,\tbar_2;\ubar_1\ubar_2}&=&\prod_{I=1,2}{1\over (T_I+\tbar_I)^2
(U_I+\ubar_I)^2}\nonumber \\ & &
\hskip -2.5cm \times\int d^2\tau\tau_2\sum_{p_L,p_R}
{p^2_L\over(T_3+\tbar_3)(U+\ubar_3)}
e^{i\pi\tau |p_L|^2}e^{i\pi{\bar\tau} |p_R|^2}
{\bar\eta}(\bar\tau)^{-2}{\rm Tr}(\delta Q^2)
\partial _{\bar\tau}\ln(\tau_2\bar{\theta}_g).
\label{F112int}
\end{eqnarray}
Here the lattice sum extends over the $\Gamma^{2,2}$ lattice of
the (untwisted)  third plane. The dependence on the moduli of the first two
planes is given by the explicit pre-factor multiplying the integral in
(\ref{F112int}) due to the normalization of the corresponding vertex operators.

In order to  evaluate the r.h.s.\ of eq.(\ref{F112int}), it is
convenient to take its derivative with respect to $\tbar_3$.
Using the identity \cite{mu}:
\begin{eqnarray}
\partial_{\tbar_3}\sum_{p_L,p_R}
{p^2_L\over(T_3+\tbar_3)(U+\ubar_3)}
e^{i\pi\tau |p_L|^2}e^{i\pi{\bar\tau} |p_R|^2}& &\nonumber\\ & &\hskip -3.5cm
={-i\over \pi\tau_2(T_3+\tbar_3)^2}\partial_{\tau}(\tau_2\partial_{U_3}
\sum_{p_L,p_R}e^{i\pi\tau |p_L|^2}e^{i\pi{\bar\tau} |p_R|^2})\ ,
\end{eqnarray}
we obtain, after a partial integration over $\tau$,
\begin{eqnarray}
\partial_{\tbar_3}F^1_{\tbar_1,\tbar_2;\ubar_1,\ubar_2}
&=&{1\over (T_3+\tbar_3)^2}\prod_{I=1,2}{1\over (T_I+\tbar_I)^2
(U_I+\ubar_I)^2}\nonumber\\ & &
\partial_{U_3}
 \int {d^2\tau\over \tau_2}\sum_{p_L,p_R}
{\bar\eta}(\bar\tau)^{-2}{\rm Tr}(\delta Q^2)
e^{i\pi\tau |p_L|^2}e^{i\pi{\bar\tau} |p_R|^2} \label{long}
\end{eqnarray}
Here we used the fact that, upon taking 
$\partial_{U_3}$ derivative, the $\tau_2\to\infty$ boundary
term vanishes for generic moduli values.
The integral on the r.h.s.\ of eq.(\ref{long}) is the familiar expression
for the group-dependent part of threshold corrections
to gauge couplings \cite{thr}, proportional to
$\ln[|\eta (iT_3)\eta (iU_3)|^4
(T_3+\tbar_3)(U_3+\ubar_3)]$. Integrating eq.(\ref{long}) we obtain
\begin{equation}
F^1_{\tbar_1,\tbar_2;\ubar_1\ubar_2} = \delta \tilde{b}\,
G_2(T_3)G_2(U_3)\prod_{I=1,2}{1\over (T_I+\tbar_I)^2
(U_I+\ubar_I)^2}\ ,  \label{short}
\end{equation}
where $\delta \tilde{b}$ is the difference between the usual
threshold function coefficients  \cite{thr} of the two gauge groups.
Note that eq.(\ref{long}) has precisely the form of the recursion relation
(\ref{hanomn}) for genus 1:
\begin{equation}
\partial_{\tbar_3}F^1_{\tbar_1,\tbar_2;\ubar_1,\ubar_2} = G^{U_3\ubar_3}
F^0_{\tbar_1,\tbar_2,\tbar_3;\ubar_1,\ubar_2,\ubar_3}\partial_{U_3}F^1
\label{rec1}
\end{equation}
with
\begin{equation}
F^1 = \delta \tilde{b}\,\sum_{I=1}^3 \ln[|\eta (iT_I)\eta (iU_I)|^4
(T_I+\tbar_I)(U_I+\ubar_I) \label{F1}
\end{equation}
and the inverse metric $G^{U_3\ubar_3}=(U_3+\ubar_3)^2$.
 
Eq. (\ref{short}) shows that the coefficient 
\begin{equation}
F^{1}_2(T_3,U_3)= \delta \tilde{b}\,G_2(T_3)G_2(U_3)
\end{equation}
of the F-term  $F^{1}_2\Pi_1\Pi_2W^{2}$ suffers from
a holomorphic anomaly analogous to the one appearing in
threshold corrections i.e.\  coefficients of the one loop $W^2$ terms.
The corresponding interactions should be in fact
represented by non-local terms in the effective action. Of course, due to the
complete symmetry among the three orbifold planes, similar anomalies occur in
all other $\Pi^2W^{2}$ terms.

At two loops, we expect holomorphic anomalies in $\Pi W^{4}$ terms.
Similarly to the one-loop case, we can take four gauginos
from the lowest component of $W^{4}$ and two anti-moduli, say
$\tbar_1$ and $\ubar_1$ from $\Pi_1$. 
Choosing a convenient kinematic
configuration, we are lead to evaluate the string amplitude:
\begin{equation}
A^{2,1}=\langle V^1_{\lambda}(x) V^2_{\lambda}(y)  V^1_{\lambda}(z) 
V^2_{\lambda}(w) 
V_{\tbar_1}(u) V_{\ubar_1}(v) \prod_{i=1}^4 e^{\varphi}T_F(z_i)\rangle\ .
\label{21}
\end{equation}
It is easy to see again by left-moving charge conservation that the vertex
operators of the two anti-moduli contribute with their fermionic parts,
providing also the two powers of momenta. The spin structure sum can be
performed by the Riemann theta-identity after a convenient gauge choice of the
position of picture changing operators along the lines of Ref. \cite{agnt}.
The result is given by the following topological amplitude:
\begin{equation}
F^2_{\tbar1;\ubar1}=\langle \dbar X_1\psibar_1(u) \dbar\xbar_1\Psi_2\Psi_3(v)
\prod_{i=1}^4 G^-(z_i)\delta (\det Q)^2\rangle
\label{F221}
\end{equation}
where $\delta (\det Q)^2$ denotes an appropriate combination of products of
determinants which ensures modular invariance, resulting from a combination of
physical amplitudes in which there is no contribution from the contraction of
the Kac-Moody currents (chosen in the Cartan subalgebra). Again, by using
left-moving charge conservation we see that two $G^-$'s provide two factors of
$\partial X_2$ and the other two  provide two factors of $\partial X_3$.
All of them are replaced by their zero-modes, given by 1-forms twisted by $g_2$
and $g_3$ respectively.\footnote{By Riemann-Roch, at genus $g$ there 
are $g-1$ twisted holomorphic one-forms, $\omega^i_h$, for the twist $h$.
Correspondingly, the bosonic zero-modes 
$\partial X=\sum_{i=1}^{g-1}p_i\omega^i_h$ for given momenta $p_i$.} 
Similarly one can see that the fields $\dbar X_1$
and $\dbar\xbar_1$ also contribute with their zero mode part, given
by the anti-holomorphic 1-form twisted by $g_1$.

Unlike in the one-loop case we are unable to compute 
$F^2_{\tbar1;\ubar1}$ directly however we will use its expected modular 
properties to write a solution of the recursion relations:
\begin{equation}
F^2_{\tbar1;\ubar1}={(\delta \tilde{b})^2\over (T_1+\tbar_1)^2(U_1+\ubar_1)^2}
\prod_{I=2,3} G_2(T_I)G_2(U_I)
\end{equation}
which indeed satisfies
\begin{equation}
\dbar_{\tbar_I}F^2_{\tbar1;\ubar1}=G^{U_I\ubar_I}
F^1_{\tbar1,\tbar_I;\ubar1,\ubar_I} \partial_{U_I}F^1
\end{equation}
for $I=2,3$ and similarly for $T_I\leftrightarrow U_I$ interchange.
We thus have
\begin{equation}
F^2_1= (\delta \tilde{b})^2\,\prod_{I=2,3} G_2(T_I)G_2(U_I)
\end{equation}
corresponding to a term 
$F^2_1\Pi^1W^4$ in the effective action.

Finally at genus 3, we have to consider a $W^6$ term with non holomorphic
coupling. The corresponding string amplitude involves two gauge bosons and
four gauginos and it was computed in section 2. The resulting topological
amplitude is given in eq.(\ref{htopf}):
\begin{equation}
F^3=\langle \prod_{i=1}^6G^-(z_i) \delta (\det Q)^2\rangle\ .
\label{F3}
\end{equation}
Since there are two fermionic zero-modes for each plane, the 6 $G^-$'s
provide 2 factors of $\dbar X_I$ for all three planes.
As in genus 2 case we have the following ansatz for $F^3$:
\begin{equation}
F^3=(\delta \tilde{b})^3\prod_{I=1}^3 G_2(T_I)G_2(U_I)
\end{equation}
which satisfies the recursion relation:
\begin{equation}
\dbar_{\tbar_I}F^3=G^{U_I\ubar_I}F^2_{\tbar_I;\ubar_I} \partial_{U_I}F^1
\end{equation}

\section{Conclusions}

In this paper we have extended the earlier results on topological
amplitudes appearing in type II strings to the case of the N=1
heterotic string. The analogues of the topological partition functions 
$F^g$'s are the amplitudes corresponding to terms of the form $W^{2g}$
in the effective action with $W$ being the gauge superfields. The main 
difference with respect to the type II case however is that the recursion 
relations do not close within such terms. We found that amplitudes
involving anti-chiral fields appear also in the recursion relations. We 
further identified these new terms as the ones corresponding to higher 
weight F-terms $W^{2g}
\Pi^n$ where $\Pi$ is a chiral superfield obtained by a chiral projection 
of a general superfield. We have also derived a set of consistent 
recursion relations obeyed by these terms in the simpler case when only
the boundary terms coming from the degeneration along dividing geodesics
contribute. This is the case for models which do not involve massless 
charged matter fields. In particular we have analyzed in detail the tree 
level terms of the form $\Pi^n$ and have given an algorithm to solve the
recursion relations based on a finite dimensional path integral. Some 
explicit orbifold examples both at the tree level as well as at higher 
loop orders were also discussed.

We have also argued that these recursion relations are just the 
supersymmetric Ward identities on the generating function of all
the connected graphs which is the object that string theory amplitudes
compute. This interpretation is strongly supported by the field theory
analysis described in the two appendices, as well as by the solution of 
the recursion relations presented in section 4.

The main open question left unanswered in this paper is the structure
of recursion relations when there are also contributions from the
handle degenerations. When there are charged massless matter fields
one would expect them to appear in the handle degeneration weighted by
their charge squares. However including such a term in the recursion relation
does not satisfy the integrability condition. A possible problem in
this could be due to the singularities in the correlation functions when 
the charged matter fields are present and interact with the gauge fields.

It is possible to write down a consistent set of recursion relations:
\begin{equation}
\theta^{\ibar}D_{\ibar} F^g = G^{j\jbar} (\sum_{g_1+g_2=g} \frac{\partial 
F^{g_1}} {\partial \eta^{\jbar}} D_j F^{g_2} + D_j \frac{\partial F^{g-1}}
{\partial \eta^{\jbar}})
\nonumber
\end{equation}
where the second term on the right hand side would be the contribution
from the handle degeneration. Notice that the second term above is not 
weighted by charge squares. We believe therefore that this equation 
describes gravitational terms of the form $W^{2g}\Pi^n$ where $W$ now is 
the gravitation chiral superfield (more precisely one should consider  
suitable differences between gravitational terms and gauge terms that 
involve no charged massless matter fields). Given our interpretation of 
the recursion relations as expressing supersymmetry Ward identity,
the second term above should be seen as quantum correction to the
latter.

The higher weight F-terms discussed in this paper are expected to play
important role for supersymmetry breaking. In particular, in the presence of
gaugino (or other fermionic) condensates induced by non perturbative effects,
they will generate a dynamical superpotential for the dilaton and moduli fields.
They could also be useful for testing duality conjectures for various $N=1$
theories \cite{N1dual}, in a similar way as $N=2$ F-terms (i.e.\ $F_g$'s) have
already provided non-trivial tests \cite{test} of type II - heterotic duality
for $N=2$ superstring compactifications \cite{N2dual}.

\vskip 0.5cm\noindent
{\bf Acknowledgments:}\\
We thank Cumrun Vafa for collaborating at the early stages of this work
and for many useful discussions. I.A. thanks ICTP and Northeastern University 
while E.G. and K.S.N. thank Ecole Polytechnique for hospitality during the
completion of this work.  
\vskip 1cm
\begin{flushleft}
{\large\bf Appendix A}\end{flushleft}
\renewcommand{\theequation}{A.\arabic{equation}}
\renewcommand{\thesection}{A.}
\setcounter{equation}{0}
 
In this appendix we illustrate the field-theoretic calculation mentioned in
section 6.  We start by adding to the standard, minimal $N=1$ supergravity
Lagrangian \cite{cremmer} a higher weight F-term of the type $\Pi^2$:
\begin{eqnarray}
L_{\Pi^2}&=& e^{-K/2}({\bar\chi}^{\bar\imath}
{\bar\chi}^{\bar\jmath}\nabla_{\bar\imath}
f_{\bar\jmath}^{(1}
- h^{\bar\imath}f_{\bar\imath}^{(1})\,(f_{{\bar k}\bar l}^{2)}\partial_m
z^{\bar k}\partial_m z^{\bar l}+f_{\bar k}^{2)}\partial^2 z^{\bar k})
\nonumber\\
&& + ~e^{-K/2}(f_{{\bar\imath}\bar\jmath}^{(1}\dslash
z^{\bar\imath}\chibar^{\bar\jmath}+
f_{\bar\imath}^{(1}\dslash\chibar^{\bar\imath})\,(f_{{\bar k}\bar l}^{2)}
\dslash z^{\bar k}\chibar^{\bar l}+f_{\bar k}^{2)}\dslash\chibar^{\bar k})
+\dots,\label{icomp}\end{eqnarray}
where a round bracket on the superscript denotes symmetrization.
As explained in section 6, in the topological theory this term
corresponds to an amplitude $F^{0}_{\ibar_1  \ibar_2;\jbar_1 \jbar_2}$
which is closed, that is $\nabla_{[\bar{k}} F^{0}_{\ibar_1  \ibar_2];\jbar_1
\jbar_2}=0$ as can be explicitly verified from (\ref{icomp}).

To perform the calculation, it is convenient to adopt a procedure which has the
advantage of keeping the correct normalization of the Ricci scalar: we introduce
Lagrange multipliers $S^{\alpha}$, $\alpha=1,2$, which are chiral superfields,
whose first component and spinor component are $s^\alpha$ and $\eta^{\alpha}$
respectively. Given the two functions $f^{\alpha}$, $\alpha=1,2$ 
specifying the $\Pi^2$ term, we introduce a modified
K\"ahler potential $\widetilde{K}=K-3\ln(1+\frac{1}{3}S_{\alpha}
f^{\alpha}+c.c.)$. Here $S_{\alpha}\equiv c_{\alpha \beta}S^{\beta}$,
where the matrix $c_{\alpha \beta}$ is defined to be equal to
1 for $\alpha\neq \beta$ and zero otherwise.
We also add to the action a superpotential term $\frac{1}{2}
S_{\alpha}S^{\alpha}=S^{1}S^{2}$. It is then clear that the equations of motion
for $S^{\alpha}$  set $S^{\alpha}=\Pi_{\alpha}$, where $\Pi_{\alpha}$  
has been defined in section 6. 

The relevant part of the supergravity Lagrangian is then:
\begin{equation}
e^{-1} L= -\frac{3}{2}[\Sigma\bar{\Sigma}e^{(-\frac{1}{3} \widetilde{K})}]_{D}+
([\frac{1}{2}{\Sigma}^{3}S_{\alpha}S^{\alpha}]_{F}+c.c.) + \dots.
\label{sugra}
\end{equation}
Here the subscripts D and F denote the action formula for a real vector 
density and for chiral density, respectively \cite{kugo}. The dots refer to
ordinary superpotential contributions. We can then solve the equations of motion
for the components $s^{\alpha}$ and $\eta^{\alpha}$ perturbatively in the number
of $f$'s and $\bar\chi$'s. Formally, we have extended the set of ordinary chiral
(anti-chiral) superfields $Z^{i}$, ($Z^{\ibar}$) by adding two more chiral
(anti-chiral) superfields $S^{\alpha}$ ($\bar S^{\bar\alpha}$). As a result, all
the geometrical quantities entering in the supergravity Lagrangian, like
metrics, connections, curvatures, being computed from $\widetilde K$, will have
components along the two additional directions. The $Z$ directions will be
denoted by latin indices, the $S$ directions with (lower case) greek indices. 

Starting from this supergravity Lagrangian, we will then compute
the full connected amplitude $\langle (\bar\chi)^4 (\bar{z})^2\rangle$
to the second order in the external momenta. We will see that there 
are two types of kinematic structures arising: $({\bar\chi}^{\ibar}, 
{\bar\chi}^{\jbar})({\bar\chi}^{\bar k}{\bar\chi}^{\bar\ell})\partial_{\mu}
z^{\bar m}\partial_{\mu}z^{\bar n}$ and 
$({\bar\chi}^{\ibar}\bar{\sigma}^{\mu\nu} 
{\bar\chi}^{\jbar})({\bar\chi}^{\bar k}{\bar\chi}^{\bar\ell})
\newline \partial_{\mu}
z^{\bar m}\partial_{\nu}z^{\bar n}$. Both correspond in the topological
theory to $F^{0}_{\ibar_1{\ibar_2}\ibar_3 ; \jbar_1{\jbar_2}\jbar_3}$,
where the $\ibar$'s and $\jbar$'s are partitions of $\ibar$, $\jbar$,
$\bar k$, $\bar\ell$, $\bar m$ and $\bar n$ which depend on the specific
kinematic configuration in a way that will be explicitly discussed later.
These amplitudes will be quartic in the $f^{\alpha}$'s.

As explained in section 6, {\it cf}.\ eq.(\ref{fphi}), a superpotential term
contributes to the ``gravitino mass'', $m_{3/2}$, i.e. to the term
in the action proportional to $\bar{\psi}_{\mu}\bar{\sigma}^{\mu\nu}
\bar{\psi}_{\nu}$. That is in our case the $\Pi^2$ F-term gives rise to:
\begin{equation}
L_{\rm mass}= e^{{\widetilde K}/2} \bar{\psi}_{\mu}\bar{\sigma}^{\mu\nu}
\bar{\psi}_{\nu}s^{1}s^{2}\ .
\label{mass}
\end{equation}
The lowest order solution of $s_{\alpha}$ equations of motion is:
\begin{equation}
s_{\alpha}^{(0)}=c_{\alpha\beta}e^{-K/2}{\bar\chi}^{\ibar}
{\bar\chi}^{\jbar}{\nabla}_{\ibar}{\nabla}_{\jbar}f^{\beta}\ ,
\label{s0}
\end{equation}
which yields the gravitino mass:
\begin{equation}
e^{-K/2}m_{3/2}^{(0)}=({\bar\chi}^{\imath}{\bar\chi}^{\jmath})
({\bar\chi}^{\bar k}{\bar\chi}^{\bar \ell})({\nabla}_{\ibar}
{\nabla}_{\jbar}f^{1}){\nabla}_{\bar k}{\nabla}_{\bar\ell}f^{2}\ .
\label{mass0}
\end{equation}

For our purposes, we need to solve for $s_{\alpha}$ to the next order,
$s_{\alpha}^{(1)}$, cubic in the $f$'s.  This is done by expanding
in components (\ref{sugra}), both from D and F terms and varying
with respect to $\bar s$. One gets:
\begin{eqnarray}
s_{\alpha}^{(1)}&=&\frac{1}{2}s_{\alpha}^{(0)}s_{\beta}^{(0)}
f^{\beta}+\frac{1}{2}c_{\alpha\beta}f^{\beta}s_{\gamma}^{(0)}
s^{{\gamma}(0)}+c_{\alpha\beta}{\chibar}^{\ibar}{\chibar}^{\jbar}
e^{-K/2}s_{\gamma}^{(0)}[-\frac{1}{3}\nabla_{\ibar}\nabla_{\jbar}
(f^{\beta}f^{\gamma})\nonumber\\ &&
+f_{\bar k}^{\beta}G^{\bar{k}\ell}
(\nabla_{\ibar}\nabla_{\jbar}f_{\ell}^{\gamma}-\frac{1}{2}K_{\ell}
\nabla_{\ibar}\nabla_{\jbar}f^{\gamma})]\ .
\label{s-1}
\end{eqnarray}
As for the spinor component $\eta_{\alpha}$, it turns out we 
do not need the next to leading term, since it cancels out
in the action. For the leading term, $\eta_{\alpha}^{(0)}$,
we get, after expanding kinetic and potential terms and varying
with respect to $\bar\eta$:
\begin{equation}
\eta_{\alpha}^{(0)}=ie^{-K/2}(f_{{\alpha}\ibar}({\chibar}^{\ibar}
{\lover{\covslash}})+
(\nabla_{\ibar}\nabla_{\jbar}
f_{\alpha}){\chibar}^{\ibar}
{\dslash}{z}^{\jbar}).
\label{eta}
\end{equation}
Here $\cal D$ is the usual covariant derivative in $({\chi}^{\imath},
{\chibar}^{\ibar})$ space. We do not need terms involving $\bar s$ and
$\bar\eta$.

The next step is to substitute the results obtained back in the action, to get
the relevant interactions. We will give some of the terms that enter into our
computation. From the expansion of the bosonic kinetic terms we have:
\begin{eqnarray}
L_{\rm bk}&=&(f_{\alpha\jbar}\Box z^{\jbar}+\nabla_{\ibar}\nabla_{\jbar}
f_{\alpha}\partial_{\mu}z^{\ibar}\partial_{\mu}z^{\jbar})
(s^{\alpha (0)}+s^{\alpha (1)})\nonumber\\ &&
-\frac{1}{6}s^{\alpha(0)}s^{\beta(0)}[(f_{\alpha}f_{\beta})_{\jbar}
\Box z^{\jbar}+\nabla_{\ibar}\nabla_{\jbar}(f_{\alpha}f_{\beta})
\partial_{\mu}z^{\ibar}\partial_{\mu}z^{\jbar}],
\label{bk}
\end{eqnarray}
where $\Box z^{\jbar}\equiv{\partial}^{2}z^{\jbar}+
{\Gamma}_{\bar{k},\ibar}^{\jbar}\partial_{\mu}z^{\bar
k}\partial_{\mu}z^{\ibar}$.

The same must be done for the fermionic part of the Lagrangian,
involving $\eta$'s and $\chi$, $\chibar$'s:
\begin{eqnarray}
L_{\rm F}&=& 2if_{\alpha\ibar}
(\chibar^{\ibar}{\lover{\covslash}})^{(0)}
{\eta^{\alpha(0)}}-2iG_{\alpha\ibar}({\chibar^{\ibar}}
{\lover{\covslash}})^{(1)}\eta^{\alpha(0)}\nonumber\\ &&
-2iG_{j\ibar}({\chibar}^{\ibar}
{\lover{\covslash}})^{(0)}\chi^{j}-
2iG_{j\ibar}({\chibar}^{\ibar}
{\lover{\covslash}})^{(1)}\chi^{j}\nonumber\\ &&
+e^{K/2}[{\eta}^{\alpha(0)}{\eta}^{\beta(0)}(e^{-s_{\alpha}^{(0)}
f^{\alpha}/2}c_{\alpha\beta}+{\widetilde K}_{\alpha}s_{\beta}^{(0)}+
{\widetilde K}_{\beta}s_{\alpha}^{(0)}-\Gamma_{\alpha\beta}^{\gamma}
s_{\gamma}^{(0)})\nonumber\\ &&
+2\eta_{\alpha}^{(0)}\chi^{i}({\widetilde K}_{i}s_{\alpha}^{(0)}
-\Gamma_{i\alpha}^{\beta}s_{\beta}^{(0)})]
+\frac{1}{2}[G_{A\bar B}G_{C\bar D}-R_{A{\bar B}C{\bar D}}]^{(0)}
\chi^{A}\chi^{C}
\chibar^{\bar B}\chibar^{\bar D}.
\label{lfermion}
\end{eqnarray}
Here the superscripts $(0)$ and $(1)$ mean that the corresponding quantities are
evaluated to zeroth and first order in $s$ and in the last term capital indices
run over both $z$ (latin) and $s$ (greek) indices. By definition,
$\chi_\alpha\equiv\eta_\alpha$.  In eqs.(\ref{lfermion}) and below
we keep track of terms up to first order in
$\chi^i$ which is sufficient for the purpose of
our calculation.

After substituting in (\ref{lfermion}) the quantities (\ref{eta}) and
(\ref{s0}), we get the desired expression for the interactions. There are two
kinds of terms, those of zeroth order in $\chi^i$, $L^{(0)}_{\rm F}$, which
give rise to irreducible contributions, and those of first order, $L^{(1)}_{\rm
F}$, giving rise to reducible diagrams.
\begin{eqnarray}
L_{\rm F}^{(0)}&=&e^{-K/2}[f_{{\alpha}\ibar}({\chibar}^{\ibar}
{\lover{\covslash}})+ (\nabla_{\ibar}\nabla_{\jbar}
f_{\alpha}){\chibar}^{\ibar}
{\dslash}{z}^{\jbar}][f_{{\beta}\jbar}({\chibar}^{\jbar}
{\lover{\covslash}})+ (\nabla_{\jbar}\nabla_{\bar k}
f_{\beta}){\chibar}^{\jbar}
{\dslash}{z}^{\bar k}]\nonumber\\ & &\hskip -1cm\times
[c_{\alpha\beta}+e^{-K/2}(\frac{1}{3}
(\nabla_{\bar m}\nabla_{\bar n}(f^{\alpha}f^{\beta})
-\frac{1}{2}f_{\bar m}^{\alpha}f_{\bar n}^{\beta}
-2f^{\alpha}\nabla_{\bar m}\nabla_{\bar n}f^{\beta}
\frac{1}{2} c_{\alpha\beta}f_{\gamma}
\nabla_{\bar m}\nabla_{\bar n}f^{\gamma})]\nonumber\\ & &\hskip -1cm
+2e^{-K}[f_{{\alpha}\ibar}({\chibar}^{\ibar}
{\lover{\covslash}})+ (\nabla_{\ibar}\nabla_{\jbar}
f_{\alpha}){\chibar}^{\ibar}
{\dslash}{z}^{\jbar}]\nonumber\\ & &\hskip -1cm\times
[(\nabla_{\bar m}\nabla_{\bar n}f_{\gamma})
\chibar^{\bar m}\chibar^{\bar n}
(\frac{1}{3}\nabla_{\ibar}(f^{\alpha}
f^{\gamma})(\chibar^{\ibar}{\lover{\covslash}})
+\frac{1}{3}\nabla_{\ibar} \nabla_{\bar\ell}(f^{\alpha}f^{\gamma})
\chibar^{\ibar}\dslash z^{\bar\ell}
-\frac{1}{4}f^{\alpha}_{\ibar}f^{\gamma}_{\bar\ell}\chibar^{\ibar}
\dslash z^{\bar\ell})\nonumber\\ & &\hskip -1cm
+\frac{1}{4}e^{K/2}f^{\alpha}_{\ibar}f^{\gamma}
\chibar^{\ibar}\dslash(e^{-K/2}\chibar^{\bar m}\chibar^{\bar n}
\nabla_{\bar m}\nabla_{\bar n}f_{\gamma})].
\label{lf0}
\end{eqnarray}
The terms of first order in $\chi^i$ (and second order in $f$), are given by:
\begin{eqnarray}
L_{\rm F}^{(1)}&=&-2ie^{-K/2}[f_{{\alpha}\jbar}({\chibar}^{\jbar}
{\lover{\covslash}})+
(\nabla_{\bar k}\nabla_{\jbar}
f_{\alpha}){\chibar}^{\jbar}
{\dslash}{z}^{\bar k})]\chi^i(\nabla_{\ibar}\nabla_{\bar\ell}
f^{\alpha}_{i}-K_{i}\nabla_{\ibar}\nabla_{\bar\ell}f^{\alpha}
-\frac{1}{2}f_{\bar\ell}^{\alpha}G_{i\ibar})
\chibar^{\ibar}\chibar{\bar\ell}\nonumber\\ & &
-2ie^{-K/2}[f_{\alpha,i\jbar}(\chibar^{\jbar}{\lover{\covslash}})+
(\chibar^{\jbar}\dslash z^{\bar k})\nabla_{\jbar}\nabla_{\bar k}f_{\alpha,i}
\nonumber\\ & &
-\frac{1}{4}G_{i\jbar}\chibar^{\jbar}(f_{\alpha,\bar k}\dslash z^{\bar k}-
f_{\alpha,k}\dslash z^{k})]\chi^{i}
(\chibar^{\bar m}\chibar^{\bar n})
\nabla_{\bar m}\nabla_{\bar n}f^{\alpha}\nonumber\\ & &
-\frac{i}{2}G_{i\jbar}{\chibar}^{\jbar}f_{\alpha}
\dslash[\nabla_{\bar k}\nabla_{\bar\ell}(e^{-K/2}f^{\alpha}){\chibar}^{\bar k}
{\chibar}^{\bar\ell}]{\chi}^{i}.
\label{lf1}
\end{eqnarray}

We are now in position to compute the amplitude $\langle{\chibar}^{4}
{\bar{z}}^{2}\rangle$, to second order in external momenta. Let us start
from amplitudes of the type $({\bar\chi}^{\ibar}
{\bar\chi}^{\jbar})({\bar\chi}^{\bar k}{\bar\chi}^{\bar\ell})\partial_{\mu}
z^{\bar m}\partial_{\mu}z^{\bar n}$. It turns out that, for this case,
(\ref{lf1}) is the only term  
we need from the fermionic sector (\ref{lfermion}).
We have, to begin with, irreducible contributions $A_{\rm irr}$  
which can be read off directly from (\ref{bk}). They are given by:
\begin{eqnarray}
2e^{K}A_{\rm irr}&=&-\frac{1}{3}(\nabla_{\bar m}\nabla_{\bar n}(f_{\alpha}
f_{\beta}))(\nabla_{\bar i}\nabla_{\bar\jmath}f^{\alpha})
\nabla_{\bar k}\nabla_{\bar\ell}f^{\beta}+
(\nabla_{\bar m}\nabla_{\bar n}
f_{\alpha})f^{\alpha}(\nabla_{\bar k}\nabla_{\bar\ell}f^{\beta})
(\nabla_{\bar\imath}\nabla_{\bar\jmath}f_{\beta})\nonumber\\&&
+(\nabla_{\bar k}\nabla_{\bar\ell}f^{\beta})f_{\beta}
(\nabla_{\bar m}\nabla_{\bar n}f_{\alpha})(\nabla_{\ibar}\nabla_{\jbar}
f^{\alpha})-\frac{2}{3}(\nabla_{\bar k}\nabla_{\bar\ell}f_{\beta})
(\nabla_{\bar m}\nabla_{\bar n}f_{\alpha})(\nabla_{\ibar}\nabla_{\jbar}
(f^{\alpha}f^{\beta}))\nonumber\\&&
+2(\nabla_{\bar k}\nabla_{\bar\ell}f_{\beta})(\nabla_{\bar m}\nabla_{\bar n}
f_{\alpha})f^{\alpha}_{\bar p}G^{i\bar p}(\nabla_{\ibar}\nabla_{\jbar}
f_{i}^{\alpha}-\frac{1}{2}K_{i}\nabla_{\ibar}\nabla_{\jbar}f^{\beta}).
\label{irr}
\end{eqnarray}
In addition to this irreducible contribution there are three 
reducible diagrams contributing to the above amplitude, as depicted
in Figure 1. 
They are due to the presence of vertices in (\ref{bk}, \ref{lf1})
and their contribution is given below.

{}From diagram 1A:
\begin{equation}
A_{1,\rm red.}=e^{-K}(\nabla_{\ibar}\nabla_{\jbar}f_{\alpha})
f^{\alpha}_{\bar p}G^{i\bar p}D_{i}
[(\nabla_{\bar m}\nabla_{\bar n}f_{\beta})(\nabla_{\bar k}
\nabla_{\bar\ell}f^{\beta})]. 
\label{a1red}
\end{equation}
 
{}From diagram 1B:
\begin{equation}
A_{2,\rm red.}=\frac{1}{2}e^{-K}f_{\alpha}
(\nabla_{\ibar}\nabla_{\jbar}f^{\alpha})
(\nabla_{\bar m}\nabla_{\bar n}f_{\beta})(\nabla_{\bar k}
\nabla_{\bar\ell}f^{\beta}).
\label{a2red}
\end{equation}

{}From diagram 1C:
\begin{equation}
A_{3,\rm red.}=\frac{1}{2}e^{-K}(\nabla_{\ibar}\nabla_{\jbar}f_{\alpha})
f^{\alpha}_{\bar p}(\nabla_{\bar k}\nabla_{\bar\ell}f_{\beta})
f^{\beta}_{\bar q}G^{i\bar p}G^{\ell \bar q}
R_{i{\bar m}\ell {\bar n}}.
\label{a3red}
\end{equation}
\fig{4.45cm}{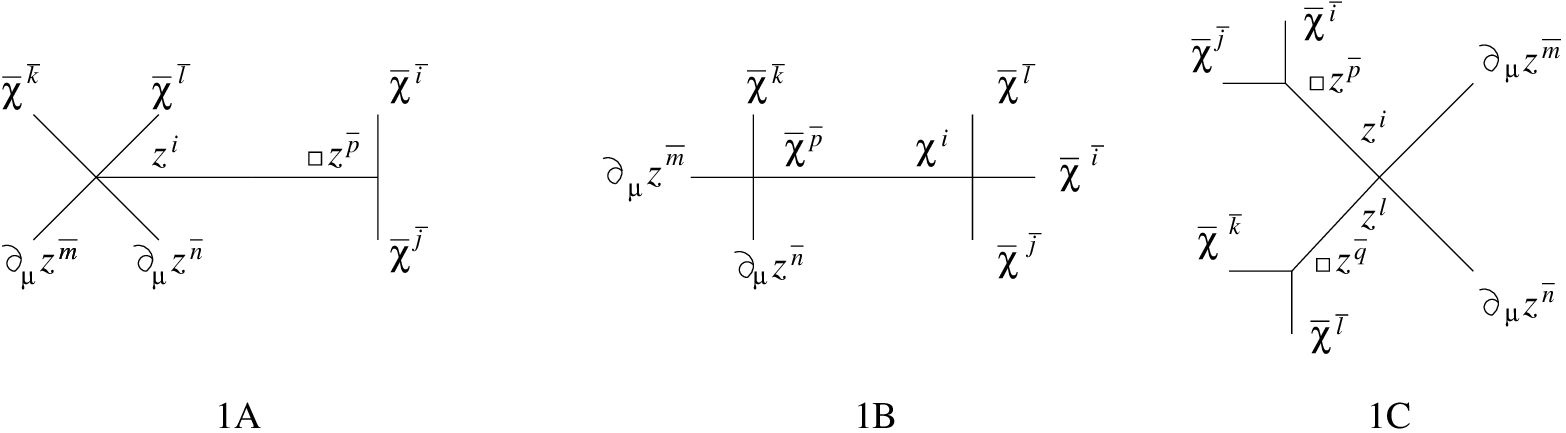}{Reducible contributions to the kinematical
structure $({\bar\chi}^{\ibar}
{\bar\chi}^{\jbar})({\bar\chi}^{\bar k}{\bar\chi}^{\bar\ell})\partial_{\mu}
z^{\bar m}\partial_{\mu}z^{\bar n}$.}{ffr}

In (\ref{a1red}) $D_i$ is the K\"ahler covariant
derivative $\partial_{i}-K_{i}$.
Adding all four contributions, we get finally for the total
amplitude $A_{\rm tot}$:
\begin{eqnarray}
2e^{K}A_{\rm tot}&=&-\frac{1}{3}(\nabla_{\bar m}\nabla_{\bar n}(f_{\alpha}
f_{\beta}))(\nabla_{\bar\imath}\nabla_{\bar\jmath}f^{\alpha})
(\nabla_{\bar k}\nabla_{\bar\ell}f^{\beta})\nonumber\\ && +
(\nabla_{\bar m}\nabla_{\bar n}
f_{\alpha})f^{\alpha}(\nabla_{\bar k}\nabla_{\bar\ell}f^{\beta})
(\nabla_{\bar\imath}\nabla_{\bar\jmath}f_{\beta})\nonumber\\ &&+
(\nabla_{\bar m}\nabla_{\bar n}f_{\alpha})
f^{\alpha}_{\bar p}G^{i\bar p}D_{i}
[(\nabla_{\ibar}\nabla_{\jbar}f_{\beta})(\nabla_{\bar k}
\nabla_{\bar\ell}f^{\beta})]\nonumber\\ &&
+R_{i{\bar m}\ell{\bar n}}
(\nabla_{\ibar}\nabla_{\jbar}f_{\alpha})
f^{\alpha}_{\bar p}(\nabla_{\bar k}\nabla_{\bar\ell}f_{\beta})
f^{\beta}_{\bar q}G^{i\bar p}G^{\ell \bar q}
+\rm {cyclic~permutations}. 
\label{tot}
\end{eqnarray}
The cyclic permutations act on the set of pairs
$({\ibar}{\jbar}),({\bar k}{\bar\ell}),({\bar m}{\bar n})$.

In a similar fashion one can study the kinematical structure
$\displaystyle ({\chibar}^{\ibar}\dslash z^{\bar m})
({\chibar}^{\jbar}\dslash z^{\bar n})({\chibar}^{\bar k}
{\chibar}^{\bar\ell})$, which contributes both to
the previous amplitude and to
$({\bar\chi}^{\ibar}\bar{\sigma}^{\mu\nu} 
{\bar\chi}^{\jbar})({\bar\chi}^{\bar k}{\bar\chi}^{\bar\ell})
\partial_{\mu}
z^{\bar m}\partial_{\nu}z^{\bar n}$, 
as it follows from the Fierz identity:
\begin{equation}
2({\chibar}^{\ibar}\dslash z^{\bar m})
({\chibar}^{\jbar}\dslash z^{\bar n})=
({\chibar}^{\ibar}{\chibar}^{\jbar})\partial_{\mu}
z^{\bar m}\partial_{\nu}z^{\bar n}+({\bar\chi}^{\ibar}\bar{\sigma}^{\mu\nu} 
{\bar\chi}^{\jbar})\partial_{\mu}z^{\bar m}\partial_{\nu}z^{\bar n}.
\label{fierz}
\end{equation}

Again one has irreducible terms as well as (four) reducible
diagrams. The irreducible contribution comes from
(\ref{lf0}). The reducible diagrams, shown in Figure 2
are obtained taking into account also the vertices of (\ref{lf1}).

Putting everything together one arrives 
at the following result for this amplitude $B_{\rm tot}$:
\fig{7.0cm}{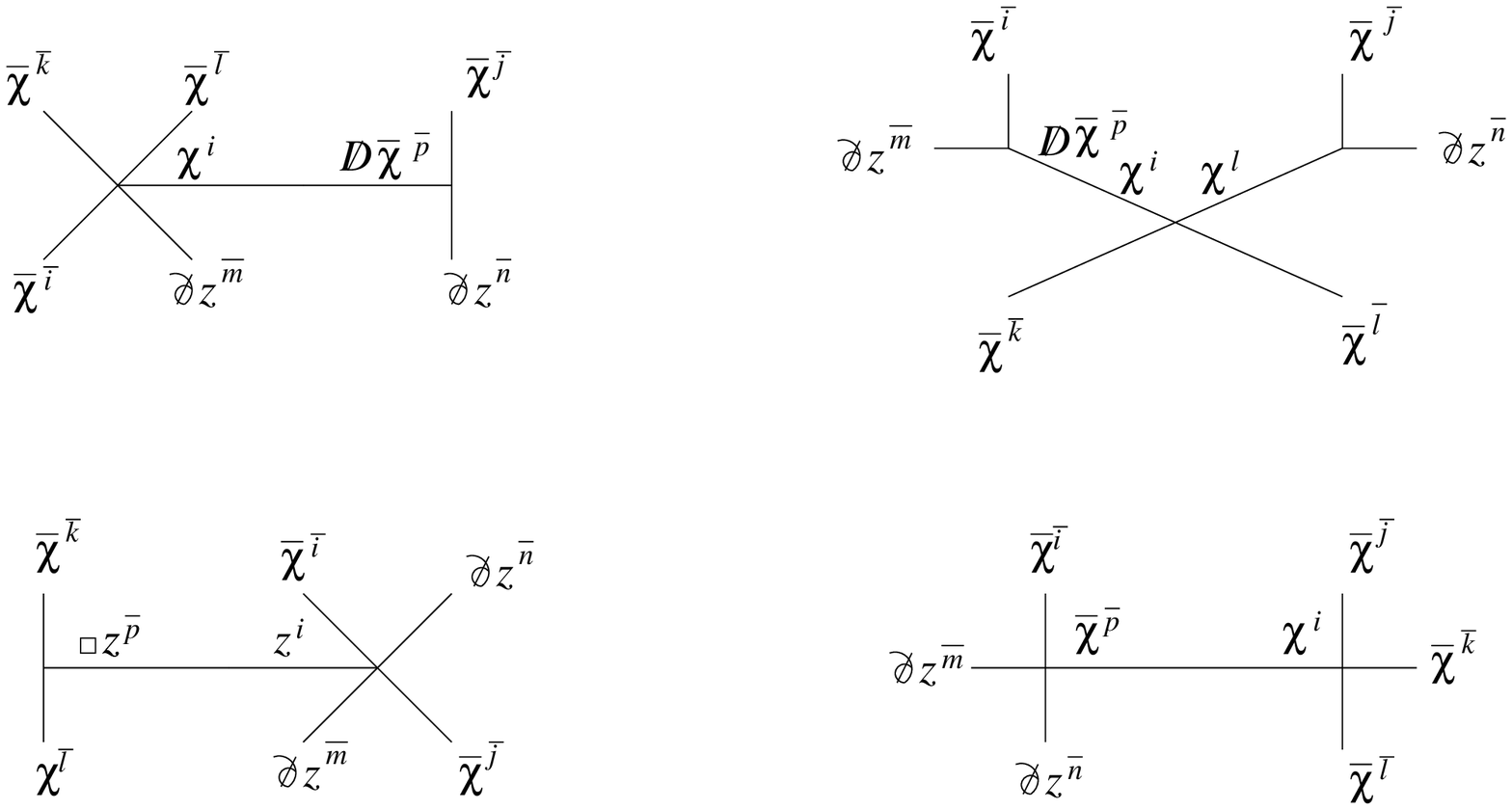}{Reducible contributions to the kinematical structure
$\displaystyle ({\chibar}^{\ibar}\dslash z^{\bar m})
({\chibar}^{\jbar}\dslash z^{\bar n})({\chibar}^{\bar k}
{\chibar}^{\bar\ell})$.}{}
\begin{eqnarray}
e^{K}B_{\rm tot}&=&-\frac{1}{3}(\nabla_{\bar k}\nabla_{\bar\ell}(f_{\alpha}
f_{\beta}))(\nabla_{\bar\imath}\nabla_{\bar m}f^{\alpha})
\nabla_{\jbar}\nabla_{\bar n}f^{\beta})\nonumber\\ && +
(\nabla_{\bar k}\nabla_{\bar\ell}
f_{\alpha})f^{\alpha}(\nabla_{\ibar}\nabla_{\bar m}f^{\beta})
(\nabla_{\bar\jmath}\nabla_{\bar n}f_{\beta})\nonumber\\ &&+
(\nabla_{\bar k}\nabla_{\bar\ell}f_{\alpha})
f^{\alpha}_{\bar p}G^{i\bar p}D_{i}
[(\nabla_{\ibar}\nabla_{\bar m}f_{\beta})(\nabla_{\jbar}
\nabla_{\bar n}f^{\beta})]\nonumber\\ &&
+R_{i{\bar k}\ell{\bar\ell}}
(\nabla_{\ibar}\nabla_{\bar m}f_{\alpha})
f^{\alpha}_{\bar p}(\nabla_{\jbar}\nabla_{\bar n}f_{\beta})
f^{\beta}_{\bar q}G^{i\bar p}G^{\ell \bar q}
+\rm {cyclic~permutations}. 
\label{btot}
\end{eqnarray}
The cyclic permutations here act on the set of pairs 
$({\ibar}{\bar m}),({\bar k}{\bar\ell})
,({\jbar}{\bar n})$.
In principle, in both cases, one could expect to have
contributions coming from exchanges
of gravitons or gravitinos. We have actually checked that
they do not contribute to the amplitudes we are considering.
  
In view of (\ref{fierz}) we can now compute the final 
expression for the two independent kinematic structures,
i.e. the one symmetric in $(\mu,\nu)$, the $\delta_{\mu\nu}$
term, coming from (\ref{tot})
and from (\ref{btot}), and that antisymmetric, containing 
$\bar{\sigma}^{\mu\nu}$, which comes from (\ref{btot}) only.
 
In either case, to make contact with the topological correlators,
one has to consider terms with a given configuration of
spinorial indices. 
Consider first the $\delta_{\mu\nu}$ term: taking, say, $\ibar$,
$\bar k$  with spinorial component $\dot{\alpha}=1$ and $\jbar$,
$\bar\ell$ with $\dot{\alpha}=2$, we see that $(\ibar,{\bar k},
{\bar n})$ as well as $(\jbar,{\bar\ell},{\bar m})$ are totally
antisymmetric. Similarly, for the $\bar{\sigma}_{\mu\nu}$
term, one can take $(\ibar,\jbar,{\bar k})$ with $\dot{\alpha}=1$
and $\bar\ell$ with $\dot\alpha=2$ and one sees that again
$(\ibar,\jbar,{\bar k})$ and $({\bar\ell},{\bar m},{\bar n})$
are totally antisymmetric.

In either case we have an amplitude  
$F^{0}_{\ibar_1{\ibar_2}\ibar_3 ; \jbar_1{\jbar_2}\jbar_3}$,
which has the form:
\begin{eqnarray}
F^{0}_{\ibar_1{\ibar_2}\ibar_3 ; \jbar_1{\jbar_2}\jbar_3}&=&
(\nabla_{\ibar_1}\nabla_{\jbar_1}f_{\alpha})f^{\alpha}_{\bar p}
G^{i \bar p}D_{i}[(\nabla_{\ibar_2}\nabla_{\jbar_2}f_{\beta})
(\nabla_{\ibar_3}\nabla_{\jbar_3}f^{\beta})]\nonumber\\ &&+
(\nabla_{\ibar_1}\nabla_{\jbar_1}f_{\alpha})f^{\alpha}_{\bar p}
(\nabla_{\ibar_2}\nabla_{\jbar_2}f_{\beta})f^{\beta}_{\bar q}
G^{\imath \bar p}G^{\ell \bar q}R_{\imath{\ibar_3}\ell{\jbar_3}}
\nonumber\\ && + (\nabla_{\ibar_1}\nabla_{\jbar_1}f_{\alpha})
f^{\alpha}(\nabla_{\ibar_2}\nabla_{\jbar_2}f_{\beta})
(\nabla_{\ibar_3}\nabla_{\jbar_3}f^{\beta})\nonumber\\ &&
-\frac{1}{3}
(\nabla_{\ibar_1}\nabla_{\jbar_1}(f^{\alpha}f^{\beta}))
(\nabla_{\ibar_2}\nabla_{\jbar_2}f_{\alpha})
(\nabla_{\ibar_3}\nabla_{\jbar_3}f_{\beta})+\dots,
\label{final}
\end{eqnarray}
where the dots indicates terms which properly antisymmetrize
$F^{0}$ in the $\ibar$'s and $\jbar$'s. Notice also that we have
removed the $e^{-K}$ factor from the string amplitudes to adhere
to the usual topological normalization, as explained in
section 2.
Now we can compute $\nabla_{[\ibar}F^{0}_{\ibar_1{\ibar_2}\ibar_3] ; 
\jbar_1{\jbar_2}\jbar_3}$. We see then that the only term which 
survives is the one where $\nabla_{\ibar}$ acts on $f^{\alpha}_{\bar p}$
in the first term of (\ref{final}) and this gives the recursion relation
(\ref{hanomn}) for the case $g=0$ and $n=3$. 
The last term in (\ref{final}) is trivially closed, whereas
the other terms cancel against those arising from the commutator
of $D_{i}$ and $\nabla_{\ibar}$.

One can follow the same procedure to compute the ``gravitino
mass'' beyond the leading order (\ref{mass0}), 
that is the connected amplitude $\langle (\chibar)^{6}
(\bar\psi)^{2}\rangle$. However in this case it turns
out that in addition to a contribution which can be
identified with 
$F^{0}_{\ibar_1{\ibar_2}\ibar_3 ; \jbar_1{\jbar_2}\jbar_3}$,
one finds also terms of different kinematic structure,
involving ratios ${p^2}/{q^2}$, where $p$ and $q$ denote generically
some intermediate momenta, whose interpretation is not clear. 
 
\begin{flushleft}
{\large\bf Appendix B}\end{flushleft}
\renewcommand{\theequation}{B.\arabic{equation}}
\renewcommand{\thesection}{B.}
\setcounter{equation}{0}

In this appendix we will work out an example illustrating
how the standard one-loop holomorphic anomaly of gauge couplings
feeds into holomorphic anomalies associated with higher-weight
interactions. More specifically, we will show that
a tree-level $\Pi^2$ term discussed in Appendix A
induces, in the presence of one-loop threshold corrections of the form
$F^1W^2$, a one-loop $\Pi W^2$ term with a coefficient $F^1_{\ibar ;\jbar}$
consistent with the recursion relations.

In order to determine $F^1_{\ibar ;\jbar}$, it is convenient to consider the
amplitude $\langle z^{\ibar}z^{\jbar}\lambda\lambda\rangle$ to second order in
the external momenta corresponding to the interaction $\partial_\mu
z^{\ibar}\partial_\mu z^{\jbar}\lambda\lambda$. We will follow the same
procedure as in the previous Appendix by introducing the Lagrange multipliers
$S^\alpha$ to describe the $\Pi^2$ interaction. The one loop threshold
corrections contain an anomalous, non-holomorphic part which
depends on both $\bar Z$ and $\bar S$. It corresponds to a non-local Lagrangian
term, however all interaction vertices involving three or more particles can be
obtained from the standard supergravity Lagrangian by taking the gauge
function $H(S,Z,\bar{S},\bar{Z})$ with the anti-chiral fields
identified as scalars, i.e.\ $\bar{Z}=\zbar$, $\bar{S}=\bar{s}$. 

In the presence of a gauge function $H\equiv F^1+S_{\alpha}H_1^{\alpha}+
{\cal O}(S^2)$, the zeroth
order equations of motion for the Lagrange multipliers yield
\begin{equation}
s_{\alpha}^{(0)}=e^{-K}c_{\alpha\beta}\lambda\lambda
(\tilde{G}_{s_{\beta}\bar{s}_{\gamma}}G^{\bar{s}_{\gamma}i}\partial_iF^1
+H_1^{\beta}) +\dots 
\label{slambda}
\end{equation}
where the matrix $\tilde{G}_{s_{\beta}\bar{s}_{\gamma}}$ is 
obtained by inverting the $2\times 2$ block $G^{s_{\beta}\bar{s}_{\gamma}}$ of
the inverse K\"ahler metric. It is straightforward to show the identity:
\begin{equation}
\tilde{G}_{s_{\beta}\bar{s}_{\gamma}}G^{\bar{s}_{\gamma}i}=
G^{i\ibar}\partial_{\ibar}f^\beta\ ,
\label{sidentity}
\end{equation}
which is valid to lowest order in $s$.
In equation (\ref{slambda}), we neglected
terms that do not contribute to the amplitude under consideration.

After substituting the solution (\ref{slambda}) into the kinetic energy
part of the Lagrangian, one finds the vertex:
\begin{equation}
(K_{s\ibar\jbar}\partial_\mu z^{\ibar}\partial_\mu z^{\jbar} + 
K_{s{\bar k}}\partial^2 z^{\bar k})s_{\alpha}^{(0)}\ .
\label{svertex}
\end{equation}
The first term contributes directly to the irreducible part of the amplitude
$\langle z^{\ibar}z^{\jbar}\lambda\lambda\rangle$, while the second term gives
rise to a reducible diagram with the scalar $z^{\bar k}$ propagating to a
3-point vertex from the kinetic terms, of the form $K_{k\ibar\jbar}z^k\partial^2
(z^{\ibar}z^{\jbar})$. As usual, the reducible diagram covariantizes the
derivatives acting on $f^\alpha$'s so that the final result becomes:
\begin{equation}
F^1_{\ibar ;\jbar}=(\nabla_{\ibar}\nabla_{\jbar}f^{\alpha})
c_{\alpha\beta}[G^{k\bar k}(\partial_{\bar k} f^\beta)(\partial_k
F^1)+H_1^{\beta}]\ ,
\label{flam}
\end{equation}
where we removed the $e^{-K}$ factor as in Appendix A. 

The expression (\ref{flam}) for $F^1_{\ibar ;\jbar}$ satisfies the identity
(\ref{ident}), $F^1_{[\ibar ;\jbar ]}=0$, in a trivial way. It also satisfies
the recursion relation (\ref{hanomn}),
\begin{equation}
\nabla_{[{\bar k}}F^1_{\ibar ];\jbar}=
F^0_{[{\bar k}\ibar ];\jbar {\bar l}}G^{l{\bar l}}\partial_l F^1\ .
\label{flamrr}
\end{equation}
After taking covariant derivative $\nabla_{\bar k}$ of (\ref{flam})
and antisymmetrizing $\ibar$ and $\bar k$ indices, one finds three terms when
the derivative acts inside the bracket. One of them gives the r.h.s. of
eq.(\ref{flamrr}) while the other two are:
\begin{equation}
(\nabla_{\jbar}\nabla_{[\ibar}f^{\alpha})
c_{\alpha\beta}[G^{l\bar l}(\partial_{\bar l} f^\beta)(\partial_{{\bar k}]}
\partial_l F^1)+\partial_{{\bar k}]}H_1^{\beta}]\ .
\label{flamder}
\end{equation}
These terms can be evaluated using the usual one loop holomorphic anomaly
equation
\begin{equation}
\partial_A\partial_{\bar B}H=bG_{A\bar B}\ ,
\label{flamanom}
\end{equation}
which is valid for gauge groups without charged matter, with $b$ proportional
to the beta-function. In our case, $A$ and $\bar B$ denote both $z$ moduli and
auxiliary $s$ fields. Using $\partial_{{\bar k}}H_1^{\beta}= G_{s_\beta {\bar
k}}=-\partial_{\bar k} f^\beta$, one sees that the two terms in
eq.(\ref{flamder}) cancel against each other, which completes the verification
of the recursion relation (\ref{flamrr}).

It is worth mentioning that the result (\ref{flam}) for $F^1_{\ibar ;\jbar}$ can
also be obtained from the gravitino ``mass" (\ref{mass}) by using the 
lowest order expressions (\ref{s0}), (\ref{slambda}).

\end{document}